\documentclass[11pt,journal,compsoc,oneside,cspaper]{IEEEtran}
\ifCLASSOPTIONcompsoc
  \usepackage[nocompress]{cite}
\else
  \usepackage{cite}
\fi

\ifCLASSINFOpdf
   \usepackage[pdftex]{graphicx}
\else
   \usepackage[dvips]{graphicx}
\fi
\usepackage{color}
\usepackage{units}
\usepackage{amsmath}
\usepackage{amssymb}
\usepackage{graphicx}

\usepackage{pifont}% http://ctan.org/pkg/pifont
\newcommand{\cmark}{\ding{52}}%
\newcommand{\xmark}{\ding{56}}%
\usepackage{array}
\ifCLASSOPTIONcompsoc
  \usepackage[caption=false,font=footnotesize,labelfont=sf,textfont=sf]{subfig}
\else
  \usepackage[caption=false,font=footnotesize]{subfig}
\fi
\usepackage[hidelinks]{hyperref}
\hypersetup{
	pdftitle={Mobile Lattice-Coded Physical-Layer Network Coding With Practical Channel Alignment},
	pdfauthor={Yihua Tan, Soung Chang Liew, and Tao Huang},
	pdfsubject={lattice-coded physical-layer network coding},
	pdfkeywords={physical-layer network coding, two-way relay network, compute-and-forward, lattice codes, channel precoding, software-defined radio},
	bookmarksnumbered=true,
	bookmarksopen=true,
	bookmarksopenlevel=1,
	colorlinks=false,
	pdfstartview=Fit,
	pdfpagemode=UseOutlines
}

\begin{document}
\title{Mobile Lattice-Coded Physical-Layer Network Coding With Practical Channel Alignment}

\author{Yihua~Tan,~%
        Soung~Chang~Liew,~\IEEEmembership{Fellow,~IEEE,}~%
        and~Tao~Huang% <-this % stops a space
\IEEEcompsocitemizethanks{\IEEEcompsocthanksitem Yihua Tan and Soung Chang Liew are with the Department of Information Engineering of the Chinese University of Hong Kong, N.T., Hong Kong.%
\IEEEcompsocthanksitem Tao Huang was a visiting research assistant at the Institute of Network Coding of the Chinese University of Hong Kong. He is with Nanjing University, China. 
}%
}% <-this % stops an unwanted space

% TODO
% funding source, paper affiliation and acknowledgement of INC

% TODO



\markboth{TAN \MakeLowercase{\textit{et al.}}: Lattice-Coded Physical-Layer Network Coding}%
{TAN \MakeLowercase{\textit{et al.}}: Lattice-Coded Physical-Layer Network Coding}


\IEEEtitleabstractindextext{%
\begin{abstract}
Physical-layer network coding (PNC) is a communications paradigm that exploits overlapped transmissions to boost the throughput of wireless relay networks. A high point of PNC research was a theoretical proof that PNC that makes use of nested lattice codes could approach the information-theoretic capacity of a two-way relay network (TWRN), where two end nodes communicate via a relay node.  The capacity cannot be achieved by conventional methods of time-division or straightforward network coding. Many practical challenges, however, remain to be addressed before the full potential of lattice-coded PNC can be realized. Two major challenges are: (1) for good performance in lattice-coded PNC, channels of simultaneously transmitting nodes must be aligned; (2) for lattice-coded PNC to be practical, the complexity of lattice encoding at the transmitters and lattice decoding at the receiver must be reduced. We address these challenges and implement a first lattice-coded PNC system on a software-defined radio (SDR) platform. Specifically, we design and implement a low-overhead channel precoding system that accurately aligns the channels of distributed nodes. In our implementation, the nodes only use low-cost temperature-compensated oscillators (TCXO)---a consequent challenge is that the channel alignment must be done more frequently and more accurately compared with the use of expensive oscillators. The low overhead and accurate channel alignment are achieved by (1) a channel precoding system implemented over FPGA to realize fast feedback of channel state information; (2) a highly-accurate carrier frequency offset (CFO) estimation method; and (3) a partial-feedback channel estimation method that significantly reduces the amount of feedback information from the receiver to the transmitters for channel precoding at the transmitters.  To reduce lattice encoding and decoding complexities, we adapt the low-density lattice code (LDLC) for use in PNC systems. Experiments show that our implemented lattice-coded PNC achieves better bit error rate performance compared with time-division and straightforward network coding systems. It also has good throughput performance in mobile non-LoS scenarios.
\end{abstract}

}

\maketitle

\IEEEdisplaynontitleabstractindextext

% \fi
\IEEEpeerreviewmaketitle

\IEEEraisesectionheading{\section{Introduction}\label{sec:intro}}

\IEEEPARstart{C}{onventional} communication systems, such as Wi-Fi, drop packets when multiple nodes transmit simultaneously. Network protocols are then devised to avoid such “collisions”. Instead of shunning collisions, a relay in a physical-layer network coding (PNC) system turns overlapping signals into a network-coded message for forwarding to their target destination nodes \cite{popovski2006anti, zhang2006hot}. The target destination nodes then extract the original message embedded within the network-coded message using self-information or side information \cite{popovski2006anti, zhang2006hot}.

\begin{figure}
	\begin{centering}
		\subfloat[\label{fig:TWRN-TD}TD.]{\begin{centering}
				\includegraphics[width=0.5\columnwidth]{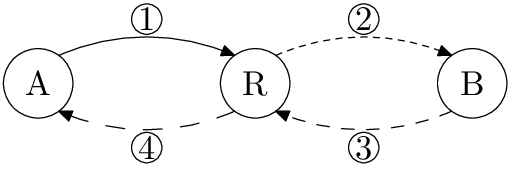}
				\par\end{centering}
		}
		\par\end{centering}
	
	\begin{centering}
		\subfloat[\label{fig:TWRN-SNC}SNC.]{\begin{centering}
				\includegraphics[width=0.5\columnwidth]{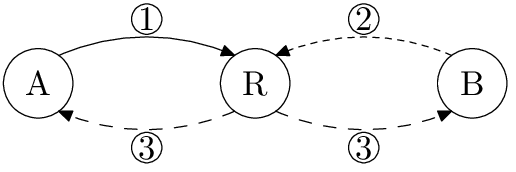}
				\par\end{centering}
		}
		\par\end{centering}
	
	\begin{centering}
		\subfloat[\label{fig:TWRN-PNC}PNC.]{\begin{centering}
				\includegraphics[width=0.5\columnwidth]{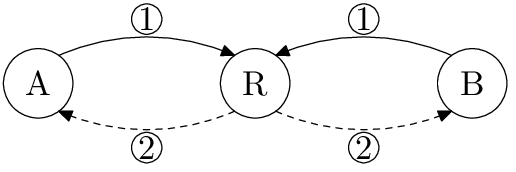}
				\par\end{centering}
		}
		\par\end{centering}
	\caption{\label{fig:TWRN}Relaying schemes in TWRN.}
\end{figure}

PNC can double the throughput of a two-way relay network (TWRN) \cite{zhang2006hot, liew2013physical}, where two end nodes A and B exchange messages with each other via a relay R. In TWRN, traditional time-division (TD) relaying requires four non-overlapping transmissions for node A to deliver a message to node B, and node B to deliver a message to node A (1. A$\rightarrow$R, 2. R$\rightarrow$B, 3. R$\leftarrow$B, 4. A$\leftarrow$R), as shown in Fig. \ref{fig:TWRN-TD}. Straightforward network coding (SNC) \cite{katti2006xors} only takes three non-overlapping transmissions by combining the two downlink transmissions into a single transmission (downlink: A$\leftarrow$R$\rightarrow$B), as shown in Fig. \ref{fig:TWRN-SNC}. Specifically, in SNC, the relay broadcasts the sum of the messages of nodes A and B. Then nodes A and B can decode the message from the other end node by subtracting their own message from the sum. PNC further reduces the required non-overlapping transmissions to two, by letting nodes A and B transmit messages simultaneously to relay R (uplink: A$\rightarrow$R$\leftarrow$B), as shown in Fig. \ref{fig:TWRN-PNC}. Relay R then decodes a network-coded message (a linear combination of the end nodes’ messages) from the overlapping signals and forward the network-coded message to nodes A and B (downlink: A$\leftarrow$R$\rightarrow$B). Upon receiving the network-coded message, each end node decodes the original message from the other end node by subtracting its own message from the network-coded message. PNC can also be applied to compute-and-forward (CF) networks \cite{nazer2011compute} where multiple source nodes transmit messages to a destination node via multiple relay nodes. Besides relay networks, PNC can also boost the performance of non-relay multiple access networks \cite{lu2014network, youncmaii2015}. 

The above TD, SNC, and PNC can be categorized as decode-and-forward (DF) relaying schemes. The traditional TD DF scheme considers one-way relaying in which the relay decodes and forwards the message from a single user, while PNC considers relay networks that involves multiple transmitters, in which the relay decodes and forwards a message combination \cite{zhang2006hot, liew2013physical, nazer2011compute, lu2014network, youncmaii2015}. SNC forwards a message combination, but unlike PNC, its relay does not receive simultaneously transmitted signals from multiple nodes.

In contrast to DF, for amplify-and-forward (AF), a relay just amplifies and forwards its received signal without decoding. AF also exploits overlapped transmissions to boost the throughput in TWRN, e.g. analog network coding (ANC) \cite{katti2007embracing}. Compared with DF, AF suffers from noise amplification and hence high error rates, especially at low signal-to-noise ratio (SNR) \cite{nosratinia2004cooperative}. Lattice-coded PNC has been proved to be able to achieve the information-theoretical capacity of TWRN to within half bit \cite{nam2010capacity}, while ANC cannot.

\textbf{Challenges and Contributions}: It has been a decade since the concept of PNC was first proposed in MobiCom 2006 \cite{zhang2006hot}. Since then, PNC has become a subfield of network coding with a wide following. Despite the many years that elapsed, most PNC research has remained theoretical. In particular, lattice codes have never been implemented in practical systems. Although some experimental PNC prototypes have been demonstrated to date \cite{lu2013implementation, chen2013frequency, lu2014network, wu2014analysis, yang2015asynchronous, youncmaii2015, kramarev2015implementation, marcum2015analysis, pan2015network, you2016reliable}, these prototypes could only achieve low-order binary phase shift keying (BPSK) and quadrature phase shift keying (QPSK) modulations. None of them makes use of lattice coding to realize the full potential of PNC. The gap between theory and implementation in the field of PNC was due to practical challenges such as channel misalignments, channel variations, and the high complexity of lattice codes. Specifically, channel misalignments and channel variations make it hard for the relay to compute a message combination with fixed integer coefficients, and the high complexity of lattice codes make it hard to be implemented. This paper presents a practical lattice-coded TWRN system that supports higher-order modulations by overcoming these challenges with a channel precoding system that accurately aligns the channels of the end nodes. In addition, the lattice encoder and decoder in the system have low complexity. Our work is overviewed below: 

\textbf{Channel alignment with low-cost commercial oscillators without reference synchronization signal}: For optimal performance of lattice-coded PNC, the channel from node A to relay R and the channel from node B to relay R must be aligned. This is to facilitate the extraction of a network-coded message from the overlapping signals at relay R. In a mobile network, the channels vary dynamically with time. Moreover, unsynchronized low-cost oscillators in general have large carrier frequency offsets (CFO) among them, which can induce a relative phase rotation of more than $2\pi$ between the overlapping signals within a packet duration. The CFO is not only large, but also keeps changing: quick and accurate estimation of the CFO is important. Some prior works dealt with these problems with reference signals, using extra antenna, extra time, and/or extra bandwidth \cite{rahul2012jmb, balan2013airsync, abari2015airshare}. We show that phase alignment can be achieved even when the nodes are driven by independent inexpensive temperature-compensated crystal oscillators (TCXO), without the need for expensive oscillators (e.g., oven-controlled crystal oscillators (OCXO) or Global Positioning System disciplined oscillators (GPSDO)), shared clocks, or a common reference frequency between nodes A and B. To enable timely channel feedback of channel state information (CSI) for precoding, we implemented the time-critical functions within the FPGA hardware of our software-defined radio (SDR) testbed. In our system, the relay estimates and feedbacks the CSI to nodes A and B with an overall feedback delay (from estimation to precoding) less than $0.5ms$. In addition, our packet format with preambles and postambles enables CFO estimation that is $100$ times more accurate than the conventional preamble-only approach in IEEE 802.11 systems.

\textbf{Time-slotted regulated transmission and proactive phase adjustment}: For orthogonal frequency-division multiplexing (OFDM) PNC, the relative phase offset of the same subcarrier of nodes A and B depend on the difference in the arrival times of their packets at the relay (i.e., the relative phase offset of a subcarrier depends on the arrival-time offset of the two packets). Prior PNC implementations \cite{lu2013implementation, lu2014network, youncmaii2015} let the relay broadcast beacons to trigger nodes A and B to transmit uplink packets together, but this beacon-trigger method cannot control the exact transmission times, causing changes to the arrival-time offset for successive uplink transmissions. This uncertainty of the arrival times will invalidate the phase precoding, because the relative phase offsets of the previous uplink transmissions may not reflect the relative phase offset of the current uplink transmissions. To remove uncertainty in arrival-time offset, we built a time-slotted system to let the end nodes transmit according to their local timers after an initial synchronization process. Then changes in arrival-time offset mainly come from sampling frequency offset (SFO), and this can be taken care of by our SFO precoding. Meanwhile, the sample shifts caused by SFO can accumulate and make the arrival-time offset be larger than the cyclic prefix (CP) of the OFDM system. We deal with this problem by letting the relay inform the lagging node to advance the sending of its packets (i.e. adjusting its time-slot boundary) once in a while. The node, besides, advancing the sending of its packets, also proactively introduces a corresponding phase precoding on each of its subcarrier to nullify the phase offset introduced by the advancement of its arrival times. In short, with our new method, we can both predict and proactively control the relative phase offset of the system. 

\textbf{Minimal overhead of signaling and feedback}: Lowering the overhead of signaling and feedback in a precoding system is very important to make the system practical. Our packet format has the same overhead as that of IEEE 802.11, incurring no extra overhead. Our precoding scheme uses a very simple protocol that does not require complicated signaling for timing synchronization and channel estimation. Moreover, our precoding scheme only requires the relay to \textit{feedback partial uplink phase and amplitude} information to the end nodes, where reciprocity is employed to construct the complete uplink phase and amplitude based on downlink measurement. We emphasize that, fundamentally, feedback cannot be totally eliminated and reciprocity will not work by itself in terms of the construction of the complete uplink phase and amplitude information (this is shown in Section \ref{sec:Channel-Alignment} of this paper). With our partial feedback scheme, the amplitude feedback overhead is cut by $98\%$, and the phase feedback overhead is cut by $96\%$ compared with the complete feedback scheme. The feedback overhead in our partial feedback scheme is negligible relative to the data payload. The phase error only increases slightly compared with the case with full feedback. 

\textbf{Practical lattice encoder and decoder}: Our system makes use of low-density lattice codes (LDLC) \cite{sommer2008low, sommer2009shaping, yona2010complex}. To our best knowledge, this is a first implementation of a lattice-coded communication system (PNC or non-PNC). For our PNC system, despite the highly accurate channel precoding, the lattice PNC decoding algorithm will still need to handle the small residual channel misalignment that varies within a packet. Such small but changing misalignment is unavoidable in practical systems. Prior theoretical studies of lattice PNC decoding did not take into account the \textit{changing} relative phase offset between the overlapping signals within a packet. We designed a novel LDLC decoder to deal with the varying residual channel misalignment. In addition, our lattice encoder and decoder incorporate lattice shaping to control the powers of lattice codewords, a must in practice due to the limited dynamic range of amplifiers. We also devised a fixed-complexity lattice shaping method to reduce computation required for shaping and decoding. 

The remainder of this paper is organized as follows: Section \ref{sec:Overview} overviews this paper, justifies the design of the lattice-coded PNC with channel precoding, and differentiates our work from related work. Section \ref{sec:Channel-Alignment} presents a practical channel precoding system that aligns the channels of the two end nodes. Section \ref{sec:LDLC} discusses LDLC encoding, shaping, and decoding (for point-to-point channel and for PNC). Section \ref{sec:sdr} presents our SDR implementation. Section \ref{sec:Experiments} presents our PNC experimental results. Section \ref{sec:discussions} discusses extension of this paper beyond TWRN. Section \ref{sec:Conclusions} concludes this paper.

\section{Overview of Main Results}\label{sec:Overview}

\begin{figure}[h]
	\begin{centering}
		\includegraphics[width=0.5\columnwidth]{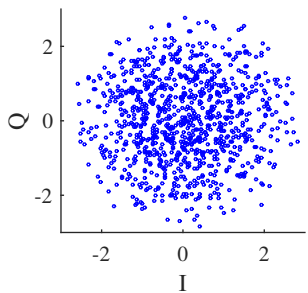}
		\par\end{centering}
	
	\vspace{-0.5em}
	\caption{\label{fig:Constellations-misaligned}Constellation of channel-misaligned
		16-QAM PNC.}
\end{figure}

This paper focuses on the uplink because for PNC performance, the uplink is the critical part---the downlink does not require channel alignment and is just ordinary multicast communication \cite{liew2013physical}. Channel alignment is very important for good performance in PNC. Channel misalignment lowers the achievable rates \cite{hern2013multilevel}. Fig. \ref{fig:Constellations-misaligned} shows the overlapped signals received by the relay when both end nodes use 16-quadrature amplitude modulation (16-QAM) over OFDM without channel alignment. The unaligned constellation makes it difficult to decode a network-coded message out of the overlapped signals. 

\begin{figure}[h]
	\begin{centering}
		\subfloat[\label{fig:ofdm-rx}Receiver-side CFO compensation and frequency-domain
		equalization in point-to-point OFDM systems.]{\begin{centering}
				\includegraphics[width=0.95\columnwidth]{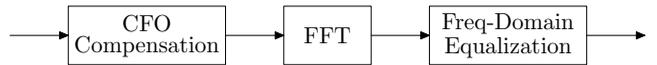}
				\par\end{centering}
		}
		\par\end{centering}
	\begin{centering}
		\subfloat[\label{fig:ofdm-tx}Transmitter-side CFO precoding and channel precoding
		in OFDM PNC.]{\begin{centering}
				\includegraphics[width=0.95\columnwidth]{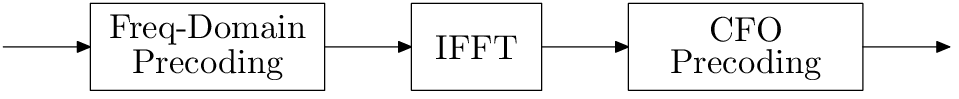}
				\par\end{centering}
		}
		\par\end{centering}
	\caption{\label{fig:ofdm-rx-tx}Receiver-side processing versus transmitter-side
		precoding.}
	
\end{figure}

In point-to-point OFDM systems, a receiver can compensate for the CFO and equalize the frequency-domain channel, as shown in Fig. \ref{fig:ofdm-rx}. But the receiver-side CFO compensation and channel equalization do not work in PNC networks, because the signals of multiple transmitters overlap at a PNC receiver. The receiver cannot do CFO compensation and channel equalization for the signals from different transmitters simultaneously. If the receiver does CFO compensation and channel equalization for one transmitter, the CFO(s) of the signals of the other transmitter(s) will not be correctly compensated for, and the channels will not be correctly equalized. Therefore, CFO compensation and channel equalization should be done at the transmitter side, rather than the receiver side. In other words, we should do channel precoding in PNC systems, where different transmitters (the two end nodes in TWRN) perform compensation and equalization according to their respective channel conditions, such that the channels of different transmitters stay aligned on each subcarrier throughout a PNC packet. As shown in Fig. \ref{fig:ofdm-tx}, the channel precoding reverses the receiver-side processing in point-to-point system: first frequency-domain channel precoding, and then time-domain CFO precoding. With the help of the transmitter-side channel precoding, the receiver does not need to do much processing before channel decoding, but it needs to estimate the channels and CFOs and feedback to the end nodes to facilitate the channel precoding at the transmitter sides. 

\begin{figure*}[t]
	\subfloat[\label{fig:Constel-OCXO-full}OCXO, full feedback]{\begin{centering}
			\includegraphics[width=0.23\textwidth]{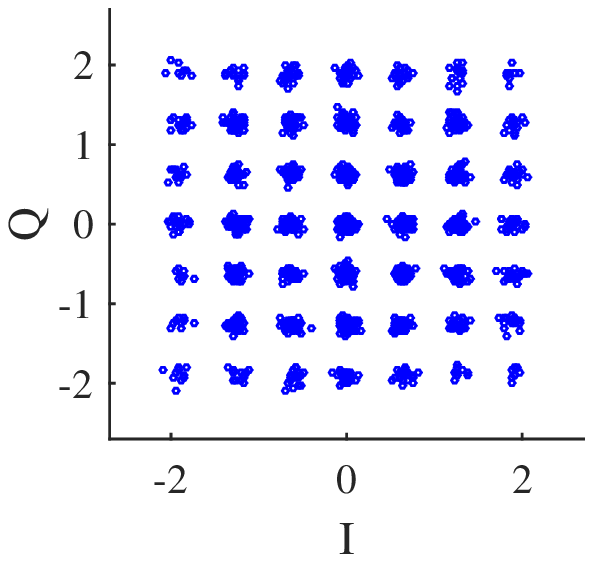}
			\par\end{centering}
		
	}\subfloat[\label{fig:Constel-OCXO-partial}OCXO, partial feedback]{\begin{centering}
			\includegraphics[width=0.23\textwidth]{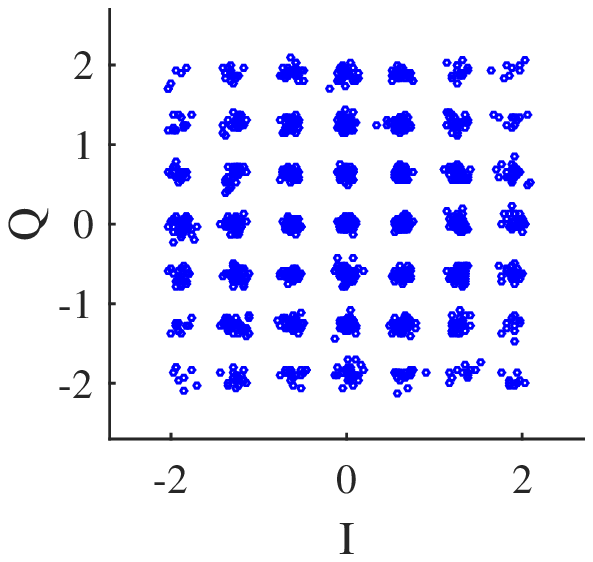}
			\par\end{centering}
		
	}\subfloat[\label{fig:Constel-TCXO-full}TCXO, full feedback]{\begin{centering}
			\includegraphics[width=0.23\textwidth]{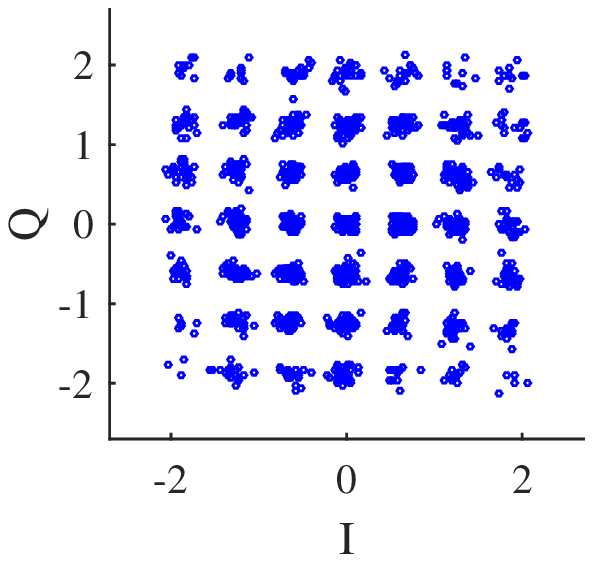}
			\par\end{centering}
		
	}\subfloat[\label{fig:Constel-TCXO-partial}TCXO, partial feedback]{\begin{centering}
			\includegraphics[width=0.23\textwidth]{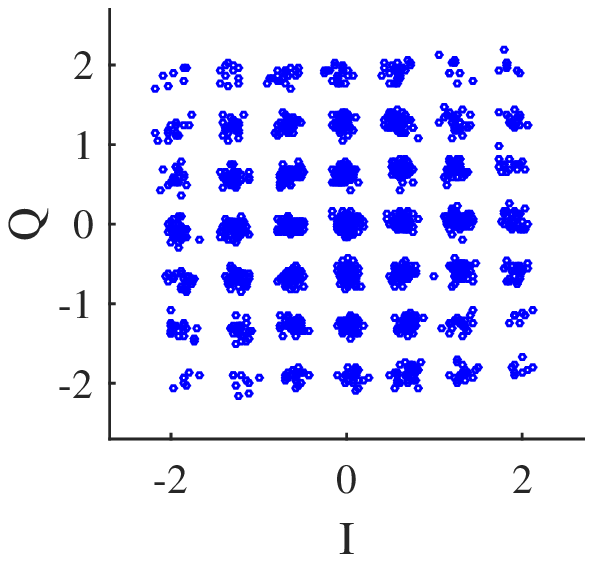}
			\par\end{centering}
		
	}
	
	\caption{\label{fig:Constellations-aligned}Constellations of channel-aligned 16-QAM PNC, with different subcarriers plotted on the same figures.}
\end{figure*}

There are many challenges to the channel precoding system, such as transmission time random jitters, feedback delays, CFO estimation errors, and feedback overhead. All these challenges are so critical that the lattice-coded PNC system will be impractical (will have poor decoding performance or high overhead) if any one of them is not correctly dealt with. Section \ref{sec:Channel-Alignment} introduces these challenges and presents our solutions in detail, and Fig. \ref{fig:System-Diagrams} in Section \ref{sec:sdr} shows the diagrams of the transmitter and receiver processing. Here we first provide some experimental results about the channel aligning performance. Fig. \ref{fig:Constellations-aligned} shows typical constellations of all subcarriers of the received signals of 16-QAM PNC, achieved using different configurations in our precoding system: 1) use OCXO (Figs. \ref{fig:Constel-OCXO-full} and \ref{fig:Constel-OCXO-partial}) or TCXO (Figs. \ref{fig:Constel-TCXO-full} and \ref{fig:Constel-TCXO-partial}); full-feedback method: feedback phases of all subcarriers (Figs. \ref{fig:Constel-OCXO-full} and \ref{fig:Constel-TCXO-full}) or partial-feedback method: only feedback the phases of 2 subcarriers out of 52 subcarriers (Figs. \ref{fig:Constel-OCXO-partial} and \ref{fig:Constel-TCXO-partial}). We use 16-QAM to illustrate our point here because the high-dimensional lattice codes actually used in our lattice-coded PNC system are hard to visualize. The $4\times 4$ 16-QAM of the two end nodes becomes $7\times 7$ 49-QAM at the relay when the overlapped signals are perfectly aligned. From Figs. \ref{fig:Constel-TCXO-full} and \ref{fig:Constel-TCXO-partial}, we see that even with inexpensive TCXO, our system can still achieve good channel alignment. From Figs. \ref{fig:Constel-OCXO-partial} and \ref{fig:Constel-TCXO-partial}, we see that our partial-feedback method only compromises the phase alignment slightly. 

\begin{figure}[h]
	\begin{centering}
		\subfloat[\label{fig:CDF}CDF of phase misalignment]{\noindent \begin{centering}
				\includegraphics[width=0.7\columnwidth]{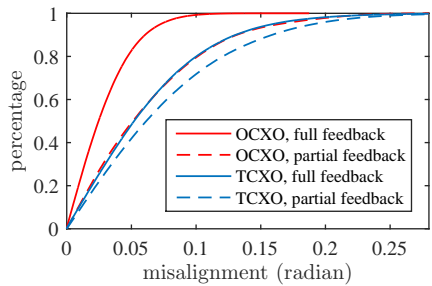}
				\par\end{centering}
		}
		\par\end{centering}
	
	\begin{centering}
		\subfloat[\label{fig:CDF-relative}CDF of phase misalignment drift]{\noindent \begin{centering}
				\includegraphics[width=0.7\columnwidth]{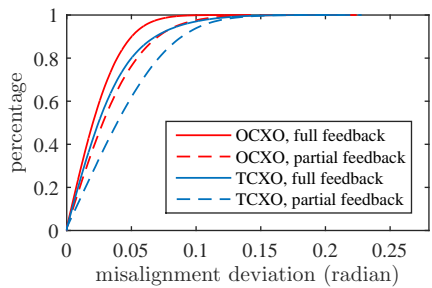}
				\par\end{centering}
		}
		\par\end{centering}
	
	\centering{}\caption{\label{fig:Empirical-CDFs}Empirical CDF of phase misalignment between the two end nodes in experiments, using different configurations.}
\end{figure}

In Fig. \ref{fig:Empirical-CDFs}, we compare the empirical cumulative distribution function (CDF) of the above four configurations. Fig. \ref{fig:CDF} plots the CDF of the phase misalignment between the two end nodes to characterize the overall phase precoding accuracy. Fig. \ref{fig:CDF-relative} plots the CDF of the phase-misalignment deviation (with respect to the misalignment of the first symbol) to characterize the misalignment drifts within a packet. These figures are obtained based on the statistics of the samples of more than $1,000$ packets, without any smoothing. 

Although using TCXO cannot achieve the same phase-alignment accuracy as using OCXO, the residual misalignment is small that we can deal with it by special lattice decoder designs. We use TCXO and partial-feedback method for the experiments in the remainder of this paper because our ultimate aim is a practical inexpensive PNC system. 

Compared with TD and SNC, although PNC requires fewer transmissions, its performance is sensitive to phase misalignment and if the phases are not aligned to a large extent, it may be able to support less dense signal constellations only---this will affect the amount of information delivered per transmission. In other words, a PNC system may potentially perform worse than TD and SNC in terms of throughput if the channels of the two end nodes are not properly aligned. Our experiment results in Section \ref{sub:ber-compare} show that with our channel precoding system, PNC can achieve lower bit error rate (BER) than TD and SNC for the same end-to-end data delivery rate (A$\rightarrow$B and A$\leftarrow$B). This experiment justifies PNC as a practical communications scheme. 

In addition to building the channel precoding system with desirable attributes (low cost, resource saving, low overhead, simple protocol, etc.), we also redesign lattice encoders and decoders to address practical considerations in PNC, including a) power control through lattice shaping, b) reduction of computation complexity caused by lattice shaping, and c) consideration of the small but varying post-precoding residual channel misalignments within a packet. Considerations a) and b) are also relevant to non-PNC systems, and by taking them into account, we have also moved the general application of lattice codes a step closer to practice. 

For traditional point-to-point communication, lattice codes allow the system to achieve the capacity of AWGN channel \cite{erez2004achieving}. However, there are also other good codes that allow the AWGN channel capacity to be achieved with lower complexity (e.g., LDPC). As a result, practical point-to-point communication systems seldom adopt lattice codes. For PNC, to our best knowledge, there has been no work showing codes other than lattice codes can allow the capacity of the two-way-relay channel to be achieved. Thus, lattice-coded PNC occupies a unique place as far as communication theory goes.

A practical challenge to lattice codes is that its decoding in general has exponential complexity. Low-density lattice codes (LDLC) \cite{sommer2008low, yona2010complex} is a new type of lattice codes with a sparse parity check matrix. LDLC can be decoded using the belief propagation (BP) algorithm in linear complexity, and systems using LDLC have been proved theoretically to be able to approach the capacity of the AWGN channel. The lattice encoding-decoding of our paper is built upon the theory in \cite{sommer2008low, yona2010complex}, with modifications to address practical issues in PNC.  Although our channel precoding scheme can align the channels to a large extent, small residual channel misalignments still remain. Furthermore, due to phase noise and tiny residual CFO and SFO, the residual channel misalignment may vary within a packet duration. To our best knowledge, all the theoretical work on lattice-coded PNC \cite{nazer2011compute, niesen2012degrees, sakzad2014phase, ordentlich2015compute} to date assume the channels to stay constant within a packet. In our system, the relay keeps track of the residual baseband channel variation using the pilot subcarriers. The relay takes into account the intra-packet channel variation when decoding a network-coded message. In addition, we use a fixed-complexity lattice shaping method to control the power of lattice codewords. This method has low shaping/decoding complexity and incurs only small performance loss compared with other high-complexity methods.

\subsection{Related Work}
The authors of \cite{kramarev2015implementation} implemented a simplified lattice-coded PNC with QPSK modulation and Reed Solomon (RS) code. However, the QPSK-RS code is still a conventional block code, and differs from lattice codes in its construction and performance. This system does not realize a lattice-coded PNC in its true spirit, and cannot approach the capacity of TWRN. Also, \cite{kramarev2015implementation} did not implement channel precoding (i.e., the channels can be misaligned, resulting in performance loss), and just used GPSDO to ensure synchronization of the oscillators. 

CF \cite{nazer2011compute} is a variant of PNC that aims to compute an integer-coefficient linear combination of the end nodes' messages.  Refs. \cite{wang2014physical, wang2015complex} analyzed the performance of CF \cite{nazer2011compute} using LDLC. Their simulation studies were limited to scenarios with perfect phase alignment, without considering how it could be achieved. Without channel alignment, the overall rate performance of the system will be far from that of channel-aligned lattice-coded PNC \cite{hern2013multilevel}. 

Channel alignment is also a problem that needs to be addressed in distributed multi-user (MU) MIMO systems. In a MU-MIMO implementation referred to as AirSync in \cite{balan2013airsync}, a master access point transmits an out-of-band reference signal to continuously calibrate the phases of slave access points. AirShare \cite{abari2015airshare} used a reference signal, but it differs from AirSync in that the reference signal is an analog clock signal. Sending continuous out-of-band reference signals requires extra bandwidth and/or antennas. In contrast, our channel precoding mechanism in PNC achieves good channel alignment without the need for extra bandwidth or antennas. 

Ref. \cite{gollakota2009interference} implemented a system of interference alignment and cancellation (IAC). In IAC, what matters is the relative phases between the multiple antennas of each client, not the absolute phase observed on each antenna. Since CFO creates the same phase rotation to the antennas of the same client, the relative phases between the antennas do not change. Therefore, CFO is immaterial as far as IAC is concerned. By contrast, what matters to PNC is the relative phase offset between the two distributed nodes driven by independent oscillators. Unlike IAC, our PNC system has to precode the CFO very accurately to maintain phase alignment.

Joint multi-user beamforming (JMB) \cite{rahul2012jmb} used an extra wireless node to transmit a reference signal to adjust the phases of different nodes in MU-MIMO.  We do not use an extra wireless node for such a purpose in our PNC system. The protocol of JMB requires rounds of uplink-downlink information exchange for each single-transmission period, incurring high signaling overhead. Our system has low overhead thanks to our simple time-slotted protocol and partial-feedback scheme. Another important difference is the assumption of JMB that CFO does not change significantly over time. We find this assumption to be invalid when TCXO is used.

Ref. \cite{shepard2012argos} presented Argos, a base station architecture in which a large number of antennas serve many terminals through multiuser beamforming. Argos exploits channel reciprocity to let the base station estimate the channels of multiple antennas relative to a reference antenna without any feedback. Although Argos performs beamforming assuming reciprocity, a fundamental difference with our work is that they used a shared clock to synchronize all the antennas on the base station. This is not practical for PNC where the nodes to be channel-aligned are located at different positions. Ref. \cite{rogalin2014scalable} applied the idea of reciprocity-based relative channel estimation to MU-MIMO to reduce the required CSI feedback in a network with multiple access points connected by a wired backhaul network. However, the assumption of wired backhaul is not valid for PNC. 

Ref. \cite{yang2013bigstation} addressed the complexity of signal processing in large-scale MIMO networks. In the future when we extend our PNC system to the many-user scenario (e.g., $N\times N$ CF, network-coded multiple access), similar consideration will be needed.

Ref. \cite{katti2007embracing} implemented ANC. ANC does not require channel alignment because the relay does not decode.  However, this approach suffers from performance loss because noise is amplified along with the signals and it cannot approach the capacity of TWRN. The ANC implementation in \cite{katti2007embracing} used minimum-shift keying (MSK) so that the end nodes can do non-coherent detection without requiring uplink channel information. But MSK is a low-order modulation, and the supportable data rate is as low as that of BPSK. Higher-order differential phase shift keying (DPSK) could be used to achieve higher data rates, while still allowing non-coherent detection. But DPSK is not power efficient when the order is higher than 4 (i.e., beyond DQPSK). On the other hand, if QAM is used for better power efficiency, then the relay needs to feedback the uplink channels (and channel variations within a packet) to the end nodes for coherent detection, inducing communication overhead. 

In Section \ref{sub:compare-precoding}, we make an overall comparison between our precoding system and the precoding systems in some of the prior work above. In section \ref{sub:prior-pnc}, we compare our PNC implementation with prior PNC implementations.  

Last but not least, we remark that the channel alignment in PNC is more challenging than that in distributed MIMO, because in PNC the constellations of multiple users are superimposed, resulting in a much denser constellation. For PNC, the dense but aligned constellation is needed to facilitate the computation of a network-coded message at the relay, while distributed MIMO only needs to null out unwanted interferences.

\section{Practical Channel Precoding System}\label{sec:Channel-Alignment}

The baseband channel that a receiver observes includes the air channel and the hardware of the transmitter and the receiver. The coherence time of the air channel is relatively long (e.g., ${>}10ms$) in indoor environments. The hardware-induced channel amplitudes are stable and can be seen as invariant, but the hardware-induced channel phases vary quickly due to the frequency offsets between the hardware of different nodes. So, we discuss the problems of amplitude precoding and phase precoding separately in the following. For each problem, we discuss how to ensure good aligning performance with low overhead. 

For amplitude precoding, the relay can feedback the amplitudes of all subcarriers to each end node. Since the hardware-induced amplitudes are almost invariant and the air channel amplitudes change slowly, the feedbacked amplitudes will be good enough to align the amplitudes of the two end nodes on all subcarriers. But full feedback of amplitudes on all subcarriers induces high communication overhead. We exploit channel reciprocity to reduce the overhead. Specifically, for each end node, the amplitudes of the downlink subcarriers are estimated and used to precode the uplink subcarriers to equalize the amplitudes of all subcarriers. In particular, the relay only needs to feedback an overall amplitude to each end node to balance their receive powers at the relay (i.e, there is no need to feedback the individual amplitudes of all uplink subcarriers to the end nodes). This greatly reduces the amount of feedback information required. 

For phase precoding, the relay can also feedback the phases of all subcarriers to each end node. In addition to the initial phase precoding in the frequency domain, each end node also needs to do CFO precoding in the time domain, because the baseband channel phases rotate quickly due to CFO. Even with CFO precoding, the residual CFO (due to CFO estimation errors) may still cause the phase errors to grow over time. Therefore, CFOs must be estimated accurately and phases must be feedbacked in a timely manner. The transmission time of the two end nodes must be synchronized as well, because changing time offsets will cause phase rotations in the frequency domain. After solving these problems to align phases well, the next question is how to reduce the overhead. We can also exploit channel reciprocity to reduce the overhead, but, unlike amplitudes, an end node cannot derive the uplink phase of a subcarrier from the downlink phase of the same subcarrier---the reason for that will be elaborated in Section \ref{sub:Difficulties-Phase}. Nevertheless, as will be demonstrated in Section \ref{sub:reciprocal-phase}, a method that exploits reciprocity to reduce phase feedback to a minimal amount is still possible. 

In the following, Section \ref{sub:Amplitude-Precoding} discusses amplitude precoding, and Sections \ref{sub:Difficulties-Phase} to \ref{sub:reciprocal-phase} discuss phase precoding. Section \ref{sub:compare-precoding} compares our precoding system with the precoding systems of prior work. 

\subsection{Reciprocity-Based Amplitude Precoding}\label{sub:Amplitude-Precoding}
To keep the amplitudes of different end nodes aligned throughout a packet, the amplitude precoding in PNC has to balance the powers of the signals of the end nodes for each and every subcarrier. To achieve overall power balance, the relay feedbacks an amplitude scaling factor to each end node to precode its average power. To equalize the amplitudes of different subcarriers, we exploit channel reciprocity and let each end node precode its uplink amplitudes according to the amplitudes of the received downlink packets. The next paragraph gives experimental results demonstrating the viability of amplitude precoding based on reciprocity.

\begin{figure}[h]
	\begin{centering}
		\includegraphics[width=0.9\columnwidth]{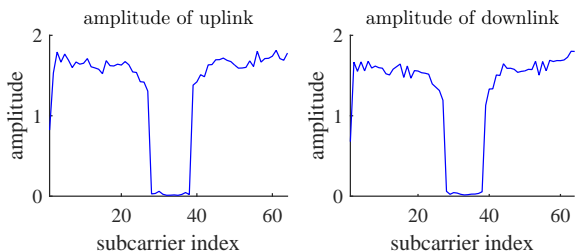}
		\par\end{centering}
	
	\vspace{-0.5em}
	\caption{\label{fig:Amplitudes-Reciprocity}Amplitudes of the uplink and
		the downlink channels.}
\end{figure}

\begin{figure}[h]
	\begin{centering}
		\includegraphics[width=0.5\columnwidth]{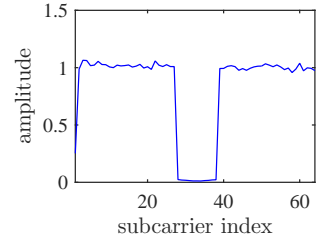}
		\par\end{centering}
	
	\vspace{-0.5em}
	\caption{\label{fig:Equalized-Amplitude}Equalized Amplitude.}
\end{figure}

Fig. \ref{fig:Amplitudes-Reciprocity} shows the amplitudes of an uplink packet and a downlink packet transmitted one after another in an experiment. Note that subcarrier $k=0$ is unused DC and subcarriers $k=27\sim 37$ are the null guard band; thus these subcarriers have near-zero amplitudes. We can see that the channel amplitudes of the uplink and the downlink are reciprocal to a large extent. The slight difference of the channel amplitudes of the uplink and downlink is due to the circuit difference of the TX paths and RX paths of nodes A and B. Because the amplitude differences induced by circuits are very stable, we perform an initial calibration to find a scaling factor $\rho_k$ for each subcarrier $k$ such that the products of $\rho_k$ and the downlink amplitudes are proportional to the uplink amplitudes. Precoding the uplink amplitudes using the downlink estimates and $\rho_k$, we can equalize all subcarriers so that their amplitudes are almost the same, as shown in Fig. \ref{fig:Equalized-Amplitude}.

\subsection{Difficulties in PNC Phase Precoding}\label{sub:Difficulties-Phase}
A most challenging aspect of the phase precoding system is that the baseband channels are not static and may vary over time, due to several factors:

1) \textbf{The changes of physical air channels}, due to mobility and environment.

2) \textbf{The large and varying CFO}, due to the use of independent RF oscillators at the two end nodes.

3) \textbf{The packet-to-packet arrival-time jitters}, due to imperfect synchronization of the packet transmission times at the two end nodes.

4) \textbf{The phase reciprocity of the uplink channel and the downlink channel does not work without regular feedback}, even with initial calibration. 

The impact of the first two factors on phase variation over time, hence how frequent phase re-estimation needs to be performed, is obvious. That of the last two factors is less so. The remainder of this subsection will elaborate on the third factor. The fourth factor will be detailed in Section \ref{sub:reciprocal-phase}.

The beacon-triggered mechanism of the prior implementations of non-precoded OFDM PNC systems \cite{lu2013implementation, lu2014network, youncmaii2015} suffers from \textit{unaccounted} packet-to-packet arrival-time jitters. In these implementations, the relay sends a beacon packet to trigger the end nodes to transmit uplink packets simultaneously. The synchronization/alignment requirement was loose in these systems: the arrival time of the later packet only needs to be within the cyclic prefix (CP) of the earlier packet \cite{liew2013physical, lu2013implementation}.  The beacon-triggered mechanism can easily meet the within-CP requirement, but a misalignment of one sample may be inevitable due to the resolution of arrival-time estimate at the end nodes. 

\begin{figure}[h]
	\begin{centering}
		\includegraphics[width=0.99\columnwidth]{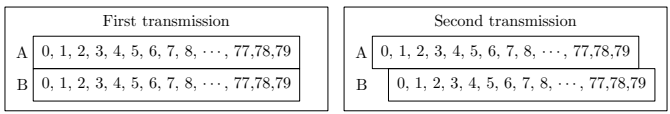}
		\par\end{centering}
	
	\caption{\label{fig:Arrival-time-jitters}Packet-to-packet arrival-time jitters.}
\end{figure}

Within-CP arrival, however, is not good enough for the precoded PNC, because varying phase offset from subcarrier to subcarrier can be induced by tiny misalignment of arrival times.  We give an example to illustrate the problem. In Fig. \ref{fig:Arrival-time-jitters}, suppose that in the first uplink transmission the packets of the two end nodes are perfectly aligned in the time domain, but in the second transmission they are separated by one sample. The relay estimates the CSI from the first transmission, and feedbacks the CSI to nodes A and B for their precoding. However, the feedback CSI is invalid for the second transmission, because the time-domain alignment is different from the last time and will effectively introduce a different set of phase offsets from subcarrier to subcarrier in the frequency domain. Note that the relay cannot compensate for this channel change by adjusting the CP-cut position. Specifically, for a PNC system, the relay cannot compensate for the channels of the two users at the same time: compensating for one inevitably leads to an uncompensated channel of the other.

We now explain the fourth factor. For simplicity, let us focus on a particular subcarrier. Denote the uplink phase the subcarrier at time $t$ by $\theta_{\mathrm{up}}\left(t\right)$, and the downlink phase by $\theta_{\mathrm{dn}}\left(t\right)$. The changes of the uplink phases and downlink phases from time $t_1$ to time $t_2$ are caused by 1) reciprocal change of physical channel, denoted by $\gamma(t_1,t_2)$, and 2) opposite change due to clock difference (CFO, SFO, and phase noise), denoted by $\eta(t_1,t_2)$ for the uplink and thus $-\eta(t_1,t_2)$ for the downlink. We therefore have 
\begin{equation}
\label{eq:uplink-phase}
\theta_{\mathrm{up}}\left(t_{2}\right)=\theta_{\mathrm{up}}\left(t_{1}\right)+\eta\left(t_{1},t_{2}\right)+\gamma\left(t_{1},t_{2}\right)
\end{equation}
and 
\begin{equation}
\label{eq:downlink-phase}
\theta_{\mathrm{dn}}\left(t_{2}\right)=\theta_{\mathrm{dn}}\left(t_{1}\right)-\eta\left(t_{1},t_{2}\right)+\gamma\left(t_{1},t_{2}\right).
\end{equation}
Suppose that the time $t_1$ is the initial calibration time, when the end node measures $\theta_{\mathrm{dn}}\left(t_{1}\right)$ and the relay feedbacks instantaneous $\theta_{\mathrm{up}}\left(t_{1}\right)$. At a later time $t_2$, a question is whether the end node can derive the uplink phase $\theta_{\mathrm{up}}\left(t_{2}\right)$  from  the initial $\theta_{\mathrm{dn}}\left(t_{1}\right)$, $\theta_{\mathrm{up}}\left(t_{1}\right)$, and the downlink phase $\theta_{\mathrm{dn}}\left(t_{2}\right)$ measured at $t_2$ without any feedback. 

It can be easily seen that this is impossible because in the two equations \eqref{eq:uplink-phase} and \eqref{eq:downlink-phase} we have three unknowns $\theta_{\mathrm{up}}\left(t_{2}\right)$, $\eta\left(t_{1},t_{2}\right)$, and $\gamma\left(t_{1},t_{2}\right)$. In other words, from the two equations, we cannot resolve $\theta_{\mathrm{up}}\left(t_{2}\right)$. Although we can use estimated CFO and SFO to estimate the cumulative phase change $\eta\left(t_{1},t_{2}\right)$, the estimation error of $\eta\left(t_{1},t_{2}\right)$ and thus the estimation error of $\theta_{\mathrm{up}}\left(t_{2}\right)$ will grow over time as $t_2$ increases, and the phase precoding will fail soon. That is, any tiny errors in the estimated CFOs and SFOs will accumulate in the estimated phase change $\eta\left(t_{1},t_{2}\right)$ over time unless we ``reset'' the phase once in a while by feedback of the uplink phase measured at the relay. 

In a nutshell, using only information from \eqref{eq:uplink-phase} and \eqref{eq:downlink-phase} does not work and we also need feedback after the initial calibration of $\theta_{\mathrm{dn}}\left(t_{1}\right)$ and $\theta_{\mathrm{up}}\left(t_{1}\right)$. Given that we have many subcarriers, each with a phase, feedback overhead may be large. Fortunately, we can use reciprocity to reduce the feedback overhead if the cumulative phase errors are due to CFO rather than changes in the physical air channel (note: the phase changes due to physical channel changes can be measured by reciprocity using the downlink information; thus no feedback is required). Specifically, we find that the phase reciprocity expressed by \eqref{eq:uplink-phase} and \eqref{eq:downlink-phase} allows the relay to feedback the phases of only two subcarriers to each end node, while ensuring the end node can recover the phases of all subcarriers. We will elaborate on how to achieve it in Section \ref{sub:reciprocal-phase}. Let us first introduce our solutions to the challenges 1) to 3) in Sections \ref{sub:Channel-Estimation} and \ref{sub:Time-Slotted-System}.

\subsection{Channel Estimation and CFO Estimation}\label{sub:Channel-Estimation}
\begin{figure}[h]
	\begin{centering}
		\includegraphics[width=0.99\columnwidth]{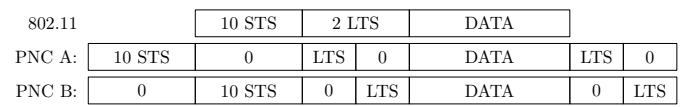}
		\par\end{centering}
	
	\caption{\label{fig:Packet-format}Packet format of 802.11, and uplink packet format of PNC.}
\end{figure}

We address challenges 1) and 2) with fast CSI feedback and accurate CFO precoding. Our implementation of lattice-coded PNC is based on OFDM, with symbol length $N=64$. As shown in Fig. \ref{fig:Packet-format}, we adopt a packet format similar to that of IEEE 802.11, but with slight modifications. The preambles consist of STS and LTS. STS is a short training sequence of $16$ samples; LTS is a long training sequence of $80$ samples (including the $16$ samples of CP). Only the DATA of the packets of nodes A and B overlap. Their STS and LTS do not overlap. We use the STS to detect packets, and the LTS before DATA to identify the starting position of the packet.

Our channel-precoded PNC system places one LTS in the preamble and one LTS in the postamble for two reasons: 1) to obtain more up-to-date channel estimation based on the postamble, and 2) to obtain much more accurate CFO estimation by correlating the preamble LTS and the postamble LTS. 

Specifically, channel estimation with preambles at the relay will induce a feedback delay equal to one PNC cycle, which consists of one uplink packet, one downlink packet, and the guard intervals in between packets. For lengthy packets, the feedback delay will then be long. Estimating the channel based on the postamble incurs a shorter feedback delay, and provides more up-to-date channel state information for the precoding of the next uplink packet.

More importantly, the packet format with preamble and postamble enables highly accurate CFO estimation, so that precoding can cancel out the CFO phase drift more completely. IEEE 802.11 systems typically estimate CFO by correlating two consecutive LTS: 
\begin{equation}
\label{eq:CFO-estimation}
\textrm{CFO}=\frac{1}{\Delta N}\textrm{angle}\left(\sum_{i=0}^{N-1}{lts}_{1}^{\dagger}\left[i\right]\cdot {lts}_{2}\left[i\right]\right)
\end{equation}
where ${lts}_{1}$ and ${lts}_{2}$ are the received signal of the two
preamble LTS, $\Delta N$ is the separation between ${lts}_{1}$ and
${lts}_{2}$, which is equal to $64$ samples in 802.11. The correlation
$\textrm{angle}\left(\sum_{i=0}^{N-1}{lts}_{1}^{\dagger}\left[i\right]\cdot {lts}_{2}\left[i\right]\right)$
finds the angle shift from ${lts}_{1}$ to ${lts}_{2}$, and the division
by their separation $\Delta N$ yields the CFO in the unit of radian/sample.

\begin{figure}[h]
	\vspace{-0.5em}
	\begin{centering}
		\includegraphics[width=0.7\columnwidth]{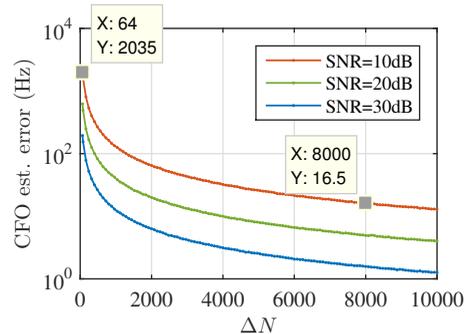}
		\par\end{centering}
	\vspace{-0.5em}
	
	\caption{\label{fig:CFO-estimation-error}CFO
		estimation error versus $\Delta N$.}
\end{figure}

The noise embedded in ${lts}_{1}$ and ${lts}_{2}$ will induce CFO estimation errors. In our system, because the preamble and the postamble are separated by the data, $\Delta N$ can be more than $8,000$ samples, much larger than the $64$-sample separation of two LTS in the 802.11 preamble. This means that the noise will be averaged over a correlation interval that is more than $100$ times longer than that of 802.11, yielding a CFO estimation $100$ times more accurate than that of 802.11. Fig. \ref{fig:CFO-estimation-error} shows CFO estimation error (in log scale) versus $\Delta N$ under different SNR. We can see that the CFO estimation error for $\Delta N=8,000$ is as small as $\nicefrac{1}{100}$ of the CFO estimation error for $\Delta N=64$.

\begin{figure}[h]

	\begin{centering}
		\includegraphics[width=0.7\columnwidth]{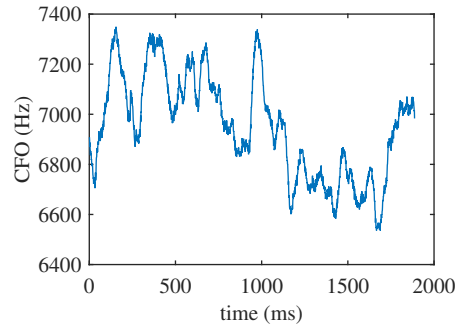}
		\par\end{centering}
	\vspace{-0.5em}
	\caption{\label{fig:CFO-variation}CFO variation.}
\end{figure}

Another idea to remove the effect of noise on CFO estimation is to average the CFO estimates over multiple packets. However, this method has the side effect of smoothing out fast CFO variations of commercial TCXO, i.e., it does not react to CFO changes quickly. Fig. \ref{fig:CFO-variation} shows the CFO estimation results for two seconds for our preambles-postamble method, with preamble-postamble distance $\Delta N=8000$ and SNR more than $20dB$, in a static environment. The CFO estimation errors are less than $10Hz$ according to Fig. \ref{fig:CFO-estimation-error}. The average CFO is about $7kHz$, but the variation can be larger than $0.4kHz$ within $100ms$. If we use the conventional 802.11 CFO estimation method and average the estimate over $100ms$ to remove noise, then the CFO estimation error can be hundreds of hertz, translating to a phase precoding error of more than $\pi/2$ radians in $1ms$. Therefore, the conventional 802.11 CFO estimation method does not work for the phase precoding in PNC.

Our preamble-postamble estimation method achieves high CFO estimation accuracy with an estimation interval (separation between the preamble and the postamble) of about 0.5 ms,  and a feedback delay (the separation between the postamble and the next packet) of about 0.5ms. With the timely CFO estimates, the end nodes can precode CFO very accurately. When testing the overall phase alignment in our system (Fig. \ref{fig:Constellations-aligned} and Fig. \ref{fig:Empirical-CDFs}), we found that smoothing CFO estimates over multiple packets did not yield any improvement when TCXO is used.  Instead, smoothing over more than 10 packets increased the phase error drastically, because the CFO coherence time is less than several milliseconds. Therefore, our reported experiments in Section \ref{sec:Experiments} just used the estimates of the CFOs and channels from the most recent packet, without smoothing. 

For simplicity of channel estimation, channel precoding and CFO precoding are performed on DATA part only, excluding the preambles and postambles. As a result, the phase drift from the preambles to the postambles can be more than one rotation as they are not CFO precoded. Meanwhile, the LTS correlation can only find the fractional rotation. In practice, we first estimate a rough CFO in the initialization phase. During the PNC experiments, we recover the full rotations based on the rough CFO and the fractional rotation, and then find the CFO. 

SFO causes sample shifts from the preambles to the postambles. In practice, the SFO is as small as ${\sim}20Hz$ for oscillators of $1ppm$ accuracy when the sampling rate (bandwidth) is $20MHz$. An SFO of $20Hz$ only induces a fractional sample shift of $20Hz\times 0.5ms=0.01$ sample in a packet of $0.5ms$ duration. The effect of such small sample shifts to the CFO estimation is negligible. 

\subsection{Time Slot Synchronization}\label{sub:Time-Slotted-System}
We address challenge 3) with tight synchronization in a time-slotted system. We built a time-slotted system to ensure two adjacent uplink packets will not have \textit{uncontrollable} changes in the alignment of packet boundaries from time slot to time slot. In the time-slotted system, nodes A and B can only transmit at the beginnings of time slots. The number of samples separating two adjacent time slots is the same for nodes A and B. Thus, if the clocks of nodes A and B were exactly synchronized, the alignment of their packets at the relay would remain the same from time slot to time slot. 

Due to the use of independent oscillators at nodes A and B, their time slots may be misaligned after a while even if they were synchronized in the beginning and that the same number of samples separate two adjacent time slots in both nodes. Thus, each end node will need to adjust its time slot boundaries once in a while to ensure time slot alignment (by momentarily adds or subtracts a few sample times from a time slot). A main difference of this mechanism with respect to the beacon-triggered mechanism is that the end nodes here know when they perform the adjustment and can compensate for the effect of the adjustment by modifying the phase of the channel precoding factor whenever it performs the time-slot adjustment (delay/advance is translated to phase change in the frequency domain at the OFDM receiver of the relay). In this way, the arrival-time jitters, which also happen in our system when an end node adjusts its time-slot boundary, can be ``accounted for'' and their effects can be compensated away through channel precoding. The next paragraph further elaborates this process. 

Note that to achieve accurate phase alignment, we do not need highly accurate alignment of the arrival times of nodes A and B. We only need to keep track of the sample shifts across adjacent time slots accurately. With our time-slotted system, the sample shifts across adjacent time slots are caused by time-slot boundary adjustment, SFO, and sampling phase noise. The end node knows the time-slot boundary adjustment exactly and compensate for the phase change induced by the adjustment. Meanwhile, SFO and sampling phase noise cause only small sub-sample shifts across adjacent time slots. SFO can be estimated accurately and compensated for by SFO precoding, and sampling phase noise has very small effects according to our experiments with TCXO. Therefore, we can align the phases without very accurate timing synchronization, such as sub-nanosecond synchronization achieved by complicated hardware design \cite{rizzi2016white}. Also, note that even an error of $0.4$ nanoseconds can result in phase change of $0.4ns\times 2.5GHz=1$ rotation. Therefore, accurate timing synchronization is neither necessary nor sufficient for accurate phase alignment. 

In the initialization phase of a PNC session, the relay sends a beacon to the end nodes to request them to start a timer. The timer defines time slots with fixed interval $T_{\mathrm{slot}}$, and all nodes transmit packets according to the time slots defined by their timers. At the beginning of each time slot, each end node transmits an uplink packet. After receiving the overlapped packets, the relay decodes the combined messages, estimates the channels of the end nodes, and then feedbacks the CSI to the end nodes. Since each end node maintains its own timer, there can be sample asynchrony between A and B.  However, the sample asynchrony remains largely the same across adjacent time slots. The constant sample asynchrony ensures that the estimated channel in time slot $1$ will still be valid for the precoding of the uplink packet in time slot $2$. 

In the long run, the SFO between the end nodes will accumulate and may cause the time slot boundaries to be offset by several samples. In our system, the relay monitors the time slot boundaries of the end nodes by correlating the received signal with the known LTS. The correlation will find $4$ peaks: in the preamble, the first correlation peak indicates the end of the LTS of end node A, and the second peak indicates the end of the LTS of end node B; likewise for the postamble. Once the relay observes that the sample asynchrony grows larger than a threshold $d_{\mathrm{thresh}}$ ($2$ samples in our experiment), it will request the lagging end node to adjust its timer ahead by $d_{\mathrm{thresh}}$ samples to catch up with the other node. This sample adjustment will change the frequency-domain channel. Thus, an end node needs to adjust the frequency-domain precoding coefficients each time it adjusts the time slot boundary. To cancel the effect of advancing by $d_{\mathrm{thresh}}$ samples, the end node multiplies the $k$-th subcarrier by 
\begin{equation}
\label{eq:Phase-adjustment}
e^{-j\cdot2\pi kd_{\mathrm{thresh}}/N}.
\end{equation}
This multiplication mimics delaying $d_{\mathrm{thresh}}$ samples in the time domain. In this way, we align slot boundaries while ensuring the channel precoding is still valid.

\begin{table*}[t]
	\caption{\label{tab:compare-precoding}Comparisons with other precoding systems. }
	
	\centering{}%
	\begin{tabular}{|c|>{\centering}p{1.45cm}|>{\centering}p{0.8cm}|>{\centering}p{0.8cm}|>{\centering}p{0.9cm}|>{\centering}p{1.7cm}|>{\centering}p{1.4cm}|>{\centering}p{2.3cm}|>{\centering}p{2.4cm}|}
		\hline 
		& Extra Antennas & Extra band & Extra node & Shared clock & Non-commercial oscillator & Feedback overhead & $95^{th}$ percentile misalignment (rad) & $95^{th}$ percentile misalignment deviation (rad)\tabularnewline
		\hline 
		This paper & \xmark & \xmark & \xmark & \xmark & \xmark & $4\%$ of full feedback & $0.18$ (partial feedback), $0.15$ (full feedback) & $0.10$ (partial feedback), $0.08$ (full feedback)\tabularnewline
		\hline 
		AirSync\cite{balan2013airsync} & \textcolor{black}{\cmark} & \textcolor{black}{\cmark} & \textcolor{black}{\cmark} & \xmark & \xmark & full & 0.08 & \tabularnewline
		\hline 
		JMB\cite{rahul2012jmb} & \xmark & \xmark & \textcolor{black}{\cmark} & \xmark & \textcolor{black}{\cmark} & full &  & 0.05\tabularnewline
		\hline 
		Argos\cite{rahul2012jmb} & \xmark & \xmark & \xmark & \textcolor{black}{\cmark} & \xmark & N/A &  & \tabularnewline
		\hline 
	\end{tabular}

\end{table*}

When implementing the time slot synchronization mechanism, the clocks of the FPGA and digital-to-analog converter (DAC) need to be configured carefully. On SDR platforms, the FPGA clock(s) and the DAC clock can be different. For example, on a WARP v3 board, the FPGA digital processing is driven by a $160MHz$ clock, while the DAC is driven by a $20MHz$ clock. Although these clocks are generated by a frequency synthesizer using a common TCXO, they may still have some relative time jitters. Therefore, if we let the FPGA run its timer (that defines the time slots) using the $160MHz$ clock, the time jitters between the two clocks will occasionally cause a one-sample jitter to the time of the DAC signal output, invalidating the phase precoding. To avoid this problem, we make the timer in sync with the DAC by configuring the FPGA to drive the timer with the same clock that drives the DAC.

\subsection{Reciprocity-Based Phase Precoding}\label{sub:reciprocal-phase}
As discussed in Section \ref{sub:Difficulties-Phase}, channel reciprocity cannot enable phase precoding without any feedback. In this subsection, we explain how to exploit channel reciprocity to minimize the amount of phase feedback information required. 

We start by subtracting \eqref{eq:uplink-phase} by \eqref{eq:downlink-phase} to get 
\begin{equation}
\theta_{\mathrm{up}}\left(t_{2}\right)=\theta_{\mathrm{up}}\left(t_{1}\right)-\theta_{\mathrm{dn}}\left(t_{1}\right)+\theta_{\mathrm{dn}}\left(t_{2}\right)+2\eta\left(t_{1},t_{2}\right).
\end{equation}
The term $\eta\left(t_{1},t_{2}\right)$ consists of the phase changes caused by CFO phase drift, SFO sample drift, carrier phase noise, and sampling phase noise. The phase change caused by CFO phase drift and carrier phase noise is common across all subcarriers. The phase change caused by SFO sample drift and sampling phase noise is linear in the shifted subcarrier index 
\begin{equation}
k^{\prime}=\begin{cases}
k, & k<\frac{N}{2}\\
k-N, & k\geq\frac{N}{2}
\end{cases}.
\end{equation}
Note that using the shifted subcarrier index, $k^{\prime}=0$ corresponds to the DC subcarrier, $k^{\prime}>0$ corresponds to a subcarrier with positive frequency, and $k^{\prime}<0$ corresponds to a subcarrier with negative frequency. Thus, the overall $\eta\left(t_{1},t_{2}\right)$ is linear in $k^{\prime}$, i.e.
\begin{equation}
\label{eq:eta-linear}
\eta\left(t_{1},t_{2}\right)=c_{0}+c_{1}k^{\prime}
\end{equation}
where $c_0,c_1\in\mathbb{R}$ are linear coefficients. It can be expanded as
\begin{equation}
\label{eq:eta-CFO-SFO}
\eta\left(t_{1},t_{2}\right)=2\pi\int_{t_{1}}^{t_{2}}\left[\mathrm{CFO}\left(\tau\right)+\mathrm{SFO}\left(\tau\right)\cdot\frac{k^{\prime}}{N}\right]d\tau
\end{equation}
if we ignore phase noise. In \eqref{eq:eta-CFO-SFO}, the CFO term $2\pi\int_{t_{1}}^{t_{2}}\mathrm{CFO}\left(\tau\right)d\tau$ is due to the cumulative phase drift caused by CFO over time. All subcarriers suffer the same phase drift due to CFO (note: before doing phase estimation using the OFDM symbol in the LTS, the receiver needs to compensate for this CFO phase drift on the OFDM symbol). The SFO induces fractional sample shifts which translate to different phase shifts in different subcarriers. In FFT, shifting integer samples $\Delta d$ causes phase rotation of $2\pi\Delta d \frac{k}{N}$, or equivalently $2\pi\Delta d \frac{k^{\prime}}{N}$ with $k^{\prime}$ as the shifted index of $k$. Although both the expressions with $k$ and $k^{\prime}$ are valid for integer-sample shifts, when the shifting theorem extends to fractional-sample shifts, the phase rotation must be expressed as proportional to the shifted index $k^{\prime}$. As the fractional-sample shift is $\int_{t_{1}}^{t_{2}}\mathrm{SFO}\left(\tau\right)d\tau$, the SFO term will be $2\pi\int_{t_{1}}^{t_{2}}\mathrm{SFO}\left(\tau\right)\cdot\frac{k^{\prime}}{N}d\tau$.

The linearity of $\eta\left(t_{1},t_{2}\right)$ suggests that the phases of any two subcarriers $k_1$ and $k_2$ should be sufficient for an end node to recover the phases of all $52$ subcarriers. Adding a superscript $k$ to $\theta_{\mathrm{up}}\left(t\right)$, we use $\theta_{\mathrm{up}}^{k}\left(t\right)$ to denote the uplink phase at subcarrier $k$, and similarly for $\eta^{k}\left(t_{1},t_{2}\right)$. From the two feedback phases $\theta_{\mathrm{up}}^{k_{1}}\left(t_{2}\right)$ and $\theta_{\mathrm{up}}^{k_{2}}\left(t_{2}\right)$, the end node can find the corresponding $\eta^{k_{1}}\left(t_{1},t_{2}\right)$ and $\eta^{k_{2}}\left(t_{1},t_{2}\right)$, then interpolate to find $\eta^{k}\left(t_{1},t_{2}\right)$ for all $k$, and finally find $\theta_{\mathrm{up}}^{k}\left(t_{2}\right)$ for all $k$. The phase precoding error does not grow with time, because in each time slot the feedback phases of two subcarriers can reset the phase error caused by estimation error of CFO and SFO.

To make the interpolation (linear regression) more robust against noise, we feedback the phases of the subcarriers with shifted index $k_1^{\prime}=-26$ and $k_2^{\prime}=26$. Two subcarriers separated this far, however, may suffer from the $2\pi$ wrap-around problem that can invalidate the phase interpolation. We resolve the $2\pi$ wrapping ambiguity by matching the interpolated phases to phases predicted by adding CFO and SFO phase drifts to the phases of the previous time slot. In addition to the above, the variation of the channel estimation times (CP cut positions of the received LTS) across successive packets and the integer-sample adjustment of time slot boundary, if not taken care of, will also mess up the phase precoding. We need to keep track of these integer sample changes in the downlink and the uplink to compensate for the accumulated integer sample differences from the initial calibration. 

Note that the above analysis is simplified by assuming the uplink phase and downlink phase can be estimated at the same time. The analysis is still valid when we consider the practical case in which the phases are estimated at different times. Just that we need to use CFO and SFO estimates to compensate for extra phase drift in the mismatched estimation times. This causes only limited phase error due to estimation errors of the CFO and the SFO, because the mismatched estimation times are short and fixed. 

Finally, with the uplink phase $\theta_{\mathrm{up}}^{k}\left(t_{2}\right)$, CFO, and SFO, an end node should precode its phases of the next uplink packet with additive phase offsets. At time $t_3$ when transmitting the next uplink packet, as in \eqref{eq:eta-CFO-SFO}, the phase change from $t_2$ to $t_3$ is $\eta^{k}\left(t_{2},t_{3}\right)=2\pi\int_{t_{2}}^{t_{3}}\left[\mathrm{CFO}\left(\tau\right)+\mathrm{SFO}\left(\tau\right)\cdot\frac{k^{\prime}}{N}\right]d\tau$, and thus the uplink phase at time $t_3$ will be $\theta_{\mathrm{up}}^{k}\left(t_{2}\right)+\eta^{k}\left(t_{2},t_{3}\right)$. To compensate for this phase change, the end node should precode its phase at time $t_3$ by precoding with
\begin{equation}
-\theta_{\mathrm{up}}^{k}\left(t_{2}\right)-2\pi\cdot\left[\mathrm{CFO}+\mathrm{SFO}\cdot\frac{k^{\prime}}{N}\right]\cdot\left[t_{3}-t_{2}\right]
\end{equation}
for subcarrier $k$. Note that $t_3$ changes within the packet. The overall error of this phase precoding process consists of the error in $\theta_{\mathrm{up}}^{k}\left(t_{2}\right)$, the estimation errors of CFO and SFO, and the channel changes from $t_2$ to $t_3$. 

Using our reciprocity-based phase precoding scheme, the relay feedbacks the phases of all subcarriers only once during the initial calibration. After that, the relay only feedbacks the phases of two subcarriers to each end node in each time slot. In each time slot, for each end node, the relay only needs to feedback $4$ coefficients: $2$ coefficients for the phases of two subcarriers, $1$ for amplitude, and $1$ for CFO. In long run, the amount of amplitude feedback information is cut by $(52-1)/52\approx 98\%$, and the amount of phase feedback information is cut by $(52-2)/52\approx 96\%$. We use a $4$-byte floating point number to represent each coefficient; so the feedback data is $16$ bytes altogether. As our packet payload length is more than $1,500$ bytes, the feedback data amount is as little as $1\%$. Note that we could further reduce the amount of feedback by quantizing a feedback coefficient with fewer bits. For the phases and amplitudes, quantization errors of $\nicefrac{1}{1000}$ are sufficient to ensure negligible penalty to the precoding. So using $10$ bits to represent a coefficient will be good enough.

\subsection{Comparisons with Other Precoding Systems}\label{sub:compare-precoding}
Tab. \ref{tab:compare-precoding} compares the resource consumptions, communication overheads, and phase alignment performances of our precoding system with those of prior work. Each of these prior precoding systems requires the use of at least one of the following: (i) extra antennas; (ii) extra band; (iii) extra node; (iv) shared clock, or (v) non-commercial oscillator. Our system does not require any of the above and is amenable to simple practical deployment. Moreover, our overall feedback overhead is just $\frac{4}{52\times 2+1}\approx 4\%$ of full feedback. AirSync \cite{balan2013airsync} and JMB \cite{rahul2012jmb} did not incorporate any methods to reduce the feedback overhead. The precoding scheme in Argos \cite{shepard2012argos} is not applicable to TWRN with distributed nodes; the overhead cannot be meaningfully compared and therefore omitted in the table. Some entries are missing in the last two columns because they are not provided in the corresponding papers. As a price for practicality, our system does incur larger phase misalignment compared with \cite{balan2013airsync, rahul2012jmb}. Fortunately, the residual phase misalignment (shown in Figs. \ref{fig:Constellations-aligned} and \ref{fig:Empirical-CDFs}) can be handled by our lattice decoder design which will be described in the next section.

\section{LDLC Encoding and Decoding}\label{sec:LDLC}

This section presents the LDLC encoding and decoding methods in our lattice-coded PNC system. Our LDLC design is built upon that in \cite{erez2004achieving, yona2010complex}, with improvement and customization on the lattice shaping and decoding to suit our PNC implementation. 

\subsection{LDLC Encoding}

A general coding lattice can be defined by a generating matrix $\mathbf{G}\in\mathbb{C}^{n\times m}$,
where $n\geq m$, with a corresponding parity check matrix $\mathbf{H}=\mathbf{G}^{-1}\in\mathbb{C}^{m\times n}$
($\mathbf{G}^{-1}$ is the pseudo inverse of $\mathbf{G}$ if $m\neq n$).
A lattice code can be defined by a coding lattice and a shaping region
to control the power of the codewords. LDLC is a special class of
lattice codes with a sparse parity check matrix $\mathbf{H}$. Each
row and each column of $\mathbf{H}$ has at most $d$ non-zeros elements.
The non-zero elements are generated from a generating sequence $\mathbf{h}=\left[h_{1},\ldots,h_{d}\right]$;
see \cite{sommer2008low,yona2010complex} for details on how
to generate $\mathbf{h}$ and $\mathbf{H}$. The sparsity of $\mathbf{H}$
allows BP decoding in linear complexity.

We encode by 
\begin{equation}
\mathbf{t}=\mathbf{G}\mathbf{b}\label{eq:LDLC_encoding}
\end{equation}
where $\mathbf{b}\in\mathbb{Z}^{m}[i]$ is a Gaussian integer vector
representing the message, and $\mathbf{t}\in\mathbb{C}^{n}$ is the
encoded lattice codeword. The encoding in \eqref{eq:LDLC_encoding}
has complexity $O(m\cdot n)$, but can be simplified by solving the
sparse equation $\mathbf{H}\mathbf{t}=\mathbf{b}$ to find $\mathbf{t}$.
We constrain each element $b_{i}$ in $\mathbf{b}$ to be a member
of the Gaussian integer alphabet 
\begin{equation}
\mathcal{B}=\left\{ b_{i,I}+jb_{i,Q},b_{i,Q}\in\mathcal{A}\right\} 
\end{equation}
where $\mathcal{A}=\{ a\in\mathbb{Z}|a_{l}\leq a\leq a_{u},\textrm{where }a_{l},a_{u}\in\mathbb{Z} \textrm{ with } $ $a_{l} \leq a_{u}\} $
is an integer alphabet.

Among all possible codewords, some codewords may have very large power.
If we just transmit $\mathbf{t}$, we will have to use a very small
gain at the RF front end (both transmitter and receiver) to avoid
saturating the amplifier. Generally, we need lattice shaping to control
the power of the lattice codeword.

Tab. \ref{tab:ldlc-notation} summarizes the notation used. In the following discussion on lattice shaping, $\tilde{\mathbf{b}}$ and $\tilde{\mathbf{t}}$ represent  the message vector and codeword vector after shaping, and $\tilde{\mathcal{A}}$ and $\tilde{\mathcal{B}}$ represent the integer alphabet and Gaussian integer alphabet of the shaped message. 

\begin{table}[h]
	\caption{\label{tab:ldlc-notation}Notation used in this section.}
	
	\begin{centering}
		
		\begin{tabular}{|c|c|}
			\hline 
			$\boldsymbol{H}$ & Sparse parity check matrix. \tabularnewline
			\hline 
			$\boldsymbol{G}$ & Lattice generating matrix. \tabularnewline
			\hline 
			$m$ & Message length. \tabularnewline
			\hline 
			$n$ & Lattice codeword length. \tabularnewline
			\hline 
			$\boldsymbol{b}$ & Message vector. \tabularnewline
			\hline 
			$\boldsymbol{t}$ & Lattice codeword vector. \tabularnewline
			\hline 
			$\mathcal{A}$ & Integer alphabet. \tabularnewline
			\hline 
			$\mathcal{B}$ & Gaussian integer alphabet. \tabularnewline
			\hline 
		\end{tabular}
		\par\end{centering}
	
\end{table}

\subsection{Fixed-Complexity Hypercube Shaping}

For a given message $\mathbf{b}$, the goal of shaping is to map $\mathbf{b}$
to $\tilde{\mathbf{b}}$ so that the resulting codeword $\tilde{\mathbf{t}}=\mathbf{G}\tilde{\mathbf{b}}$
has small power. 
By choosing a $\tilde{\mathbf{b}}$ such that $\tilde{\mathbf{b}}\equiv\mathbf{b}$ up
to a modulo operation, the receiver can obtain the original message
$\mathbf{b}$ by taking modulo after decoding $\tilde{\mathbf{b}}$
from $\tilde{\mathbf{t}}$.

We adopt hypercube shaping \cite{sommer2009shaping} in this paper.
Hypercube shaping performs shaping on $\mathbf{t}$ and $\mathbf{b}$
element-wise with the power of each element of $\tilde{\mathbf{t}}$
being individually constrained. 
Specifically, we do RQ decomposition
$\mathbf{H}=\mathbf{R}\mathbf{Q}$, where $\mathbf{R}\in\mathbb{C}^{m\times n}$
is a lower-triangular matrix, and $\mathbf{Q}\in\mathbb{C}^{n\times n}$
is a unitary matrix. 
Then from $\mathbf{Ht}=\mathbf{b}$, we have
$\mathbf{RQt}=\mathbf{R}\mathbf{t}^{\prime}=\mathbf{b}$. 
Correspondingly, we have $\mathbf{R}\mathbf{Q}\tilde{\mathbf{t}}=\mathbf{R}\tilde{\mathbf{t}}^{\prime}=\tilde{\mathbf{b}}$.
Then we perform shaping on $\mathbf{t}^{\prime}$ (i.e. obtain $\tilde{\mathbf{t}}^{\prime}$) by mapping $b_{i}$
to $\tilde{b}_{i}$ from $i=1$ to $i=m$ in a one-by-one manner.
We can determine $\tilde{b}_{i},i=1,\ldots,m$ successively thanks to the
fact that $\mathbf{R}$ is lower-triangular. 
Starting with $i=1$, for each successive $b_{i}$, we
find $\tilde{b}_{i}\equiv b_{i}$ such that the corresponding $\tilde{t}_{i}^{\prime}$
has the lowest power. The powers of $\tilde{\mathbf{t}}^{\prime}=\mathbf{Q}\tilde{\mathbf{t}}$
and $\tilde{\mathbf{t}}$ are equal because $\mathbf{Q}$ is a unitary
matrix. 
After the shaping process, the transmitter transmits $\tilde{\mathbf{t}}$.

\begin{figure}[h]
	\vspace{-1em}
	\begin{centering}
		\includegraphics[width=0.6\columnwidth]{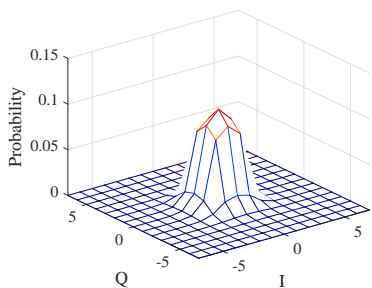} 
		\par\end{centering}
	\vspace{-0.5em}
	
	\caption{\label{fig:shaped-distribution}Probability distribution of the shaped
		Gaussian integer alphabet, where $\mathcal{A}=\left\{ -1,0,1\right\} $. }
\end{figure}

As a result of the hypercube shaping, the elements in $\tilde{\mathbf{b}}$
are no longer constrained in the alphabet $\mathcal{B}$. Rather,
the Gaussian integer alphabet of the shaped message $\tilde{\mathbf{b}}$
is 
\begin{equation}
\tilde{\mathcal{B}}=\left\{ \tilde{b}_{i,I}+j\tilde{b}_{i,Q}\vert\tilde{b}_{i,I},\tilde{b}_{i,Q}\in\tilde{\mathcal{A}}\right\} .
\end{equation}
where $\tilde{\mathcal{A}}\subset\mathbb{Z}$ is the integer alphabet
after shaping. The conventional hypercube shaping does not limit the
size of $\tilde{\mathcal{A}}$. However, the computation complexities
of the shaping and the BP decoding \cite{yona2010complex} with respect
to $|\tilde{\mathcal{A}}|$ are $O(|\tilde{\mathcal{A}}|^{2})$.
The nonlinear complexity will increase the complexity of BP decoding
dramatically if $|\tilde{\mathcal{A}}|$ is large. 
Fortunately, we find that the probability of the shaped elements $\tilde{b}_{i,I}$
or $\tilde{b}_{i,Q}$ straying far away from the origin is very low in
general. Therefore, for our implementation, we use a fixed-complexity
hypercube shaping algorithm that limits $\tilde{\mathcal{A}}$ to
\begin{equation}
\tilde{\mathcal{A}}=\left\{ \tilde{a}\in\mathbb{Z}\vert\min(\mathcal{A})-\left|\mathcal{A}\right|\leq\tilde{a}\leq\max(\mathcal{A})+\left|\mathcal{A}\right|\right\} \label{eq:integer-alphabet}
\end{equation}
i.e., the shaped integer alphabet $\tilde{\mathcal{A}}$ is triple
the size of $\mathcal{A}$, also centered around the origin. 
Limiting the range of $\tilde{\mathcal{A}}$ as such will cause the power of the resulting codeword
to be larger than ideal shaping without limits. 
To examine the penalty incurred, we did a Monte-Carlo
simulation to find the probability distribution of the shaped elements
$\tilde{b}_{i}$, where we expanded $\tilde{\mathcal{A}}$ to $\tilde{\mathcal{A}}^{\prime}=\{ \tilde{a}\in\mathbb{Z}|\min(\mathcal{A})-2|\mathcal{A}|\leq\tilde{a}\leq\max(\mathcal{A})+2|\mathcal{A}|\} $,
with $\mathcal{A}=\left\{ -1,0,1\right\} $. The probability distribution
of $\tilde{b}_{i}$ from these simulations based on $\tilde{\mathcal{A}}^{\prime}$
are shown in Fig. \ref{fig:shaped-distribution}. We observed that
the probability of $\tilde{b}_{i,I}$ or $\tilde{b}_{i,Q}$ falling
outside of the $\tilde{\mathcal{A}}$ in \eqref{eq:integer-alphabet}
is in the ballpark of $10^{-6}$. This implies that our fixed-complexity
hypercube shaping based on $\tilde{\mathcal{A}}$ in \eqref{eq:integer-alphabet}
will not have much penalty.

In the following, we first introduce the LDLC decoding for point-to-point
channel, and then extend the point-to-point decoder to our PNC decoder.

\subsection{LDLC Decoding in Single-User Systems}

After the LDLC encoding, we place the elements of the codeword $\tilde{\mathbf{t}}$
on the OFDM subcarriers, and then do Inverse Fast Fourier Transform
(IFFT) to convert the frequency-domain signal to time-domain signal $\mathbf{x}\in\mathbb{C}^{n}$
before transmission. The receiver transforms the signal back
to the frequency domain. The frequency-domain
model can be written as $\mathbf{y}=\tilde{\mathbf{t}}+\mathbf{w}$
where $\mathbf{y}\in\mathbb{C}^{n}$ is the signal after channel compensation,
and $\mathbf{w}\in\mathbb{C}^{n}$ is a noise term. The BP decoder
passes messages between the check nodes and variable codes iteratively
to decode
for $\tilde{\mathbf{b}}$. Specifically, the BP decoder checks 
\begin{equation}
\mathbf{H}\tilde{\mathbf{t}}=\tilde{\mathbf{b}}\in\tilde{\mathcal{B}}^{m}.
\end{equation}
An individual check node $i$ has a check equation given by $\sum_{k}h_{ik}\tilde{t}_{k}=\tilde{b}_{i}\in\tilde{\mathcal{B}}$,
where $k$ traverses the indices of the non-zero elements of the $i$-th
row of $\mathbf{H}$.

\subsection{LDLC Decoding in PNC}

This subsection extends the LDLC decoder to decode a linear combination
of the messages of two end nodes for the purpose of PNC. The prior theoretical works
on lattice decoding in PNC and CF \cite{nazer2011compute,wang2015complex}
proposed to decode an integer combination of the messages in the presence
of channel misalignment. Computing integer combination requires multiplying
the received signal by a scaling factor to match the integer coefficients.
The scaling operation will amplify noise and thus degrade the decoding
performance. If the phase misalignment is severe and the scaling factor
is large, then this scaling-computing scheme only works under high
signal-to-noise ratio (SNR). To achieve the best
decoding performance and avoid wasting power, it is desirable to minimize
the channel misalignment such that we can choose the scaling factor
to be small. Luckily, the phase alignment mechanism in the preceding
sections can fulfill this need, so that our
system can work well with moderate SNR. 

The prior work \cite{nazer2011compute,wang2015complex} assumed constant
random channel misalignments between users throughout a packet. The
practical case in our system is different: the phase misalignment
is small, but it varies throughout a packet. We can keep track of
the phase with the pilots in OFDM symbols (each end node has two pilots
and the pilots from different end nodes do not overlap). Since we
have precoded the channel and made the misalignment nearly zero, the
difference between different subcarriers are nearly zero. The phases
of different subcarriers may shift in the same direction due to residual CFO caused by CFO
estimation error. Therefore, for each end node, the relay can find
the average phase of its two pilots in an OFDM symbol to estimate 
the phases of all subcarriers.

In the remainder of this subsection, we introduce our LDLC decoder
that takes into account the per-symbol phase misalignments.
We aim to compute $\tilde{\mathbf{t}}_{A}+\tilde{\mathbf{t}}_{B}$. 
The frequency-domain
PNC channel model of our system is 
\begin{equation}
\mathbf{y}=\tilde{\mathbf{t}}_{A}+\mathbf{P}\tilde{\mathbf{t}}_{B}+\mathbf{w}=\tilde{\mathbf{t}}_{P}+\mathbf{w}
\end{equation}
where $\tilde{\mathbf{t}}_{A}\in\mathbb{C}^{n}$ and $\tilde{\mathbf{t}}_{B}\in\mathbb{C}^{n}$
are the lattice codewords from end nodes A and B respectively, $\tilde{\mathbf{t}}_{P}=\tilde{\mathbf{t}}_{A}+\mathbf{P}\tilde{\mathbf{t}}_{B}$
is the received combination of $\tilde{\mathbf{t}}_{A}$ and $\tilde{\mathbf{t}}_{B}$,
and $\mathbf{P}=\textrm{diag}\left(p_{1},\ldots,p_{n}\right)$ represents
the phase offset between $\tilde{\mathbf{t}}_{A}$ and $\tilde{\mathbf{t}}_{B}$,
with $p_{j}$ representing the channel misalignment (in both amplitude
and phase) in the $j$-th dimension (estimated with the pilots). Note
that we have normalized the channel of A to be one. In general, $p_{1},\ldots,p_{n}$
are different. We scale $\mathbf{y}$ element by element separately
with $\beta_{1},\ldots,\beta_{n}$ to obtain $\tilde{y}_{j}=\beta_{j}y_{j}$.
In a matrix form, we have $\tilde{\mathbf{y}}=\mathbf{B}\mathbf{y}$ where
$\mathbf{B}=\textrm{diag}(\beta_{1},\ldots,\beta_{n})$. The goal
of the scaling is to match 
\begin{equation}
\tilde{y}_{j}=\beta_{j}\tilde{t}_{A,j}+\beta_{j}p_{j}\tilde{t}_{B,j}+\beta_{j}w_{j}
\end{equation}
to $\tilde{t}_{A,j}+\tilde{t}_{B,j}$ better. By minimal mean square
error (MMSE) rule, we have 
\begin{equation}
\beta_{j}=\frac{1+p_{j}^{\dagger}}{\sigma^{2}+1+|p_{j}|^{2}}
\end{equation}
where $\sigma^{2}$ is the noise variance. After the scaling operation,
we use the LDLC decoder in the last subsection to decode $\tilde{\mathbf{b}}_{A}+\tilde{\mathbf{b}}_{B}$.
Finally, we calculate 
\begin{equation}
\mathbf{b}_{\textrm{PNC}}=(\tilde{\mathbf{b}}_{A}+\tilde{\mathbf{b}}_{B})\bmod\mathcal{A}=\left(\mathbf{b}_{A}+\mathbf{b}_{B}\right)\bmod\mathcal{A}.
\end{equation}
The relay then sends $\mathbf{b}_{\textrm{PNC}}$ to the two end nodes.
Each end node can recover the message from the other end node by subtracting
its own message from $\mathbf{b}_{\textrm{PNC}}$ and then taking
modulo on $\mathcal{A}$.

\textbf{The Effect of Shaping on Decoding}: The conventional LDLC decoders \cite{sommer2008low,yona2010complex,wang2015complex} assumed equiprobable alphabet, and extended the check node messages with respect to all elements in the alphabet in the same way in the BP iterations. However, Fig. \ref{fig:shaped-distribution} shows that the hypercube shaping yields an alphabet with Gaussian-like probability distribution, and thus the probability distribution of the checksum is also Gaussian-like. We also take into account this more accurate probability distribution in our decoder. In the message passing between the check nodes and variable nodes in the BP iterations, we scale the information from a check sum ($\tilde{b}_{A,j}+\tilde{b}_{B,j}$) with a weight proportional to its probability.

\section{SDR Implementation}\label{sec:sdr}
\begin{figure}[h]
	\begin{centering}
		\includegraphics[width=0.8\columnwidth]{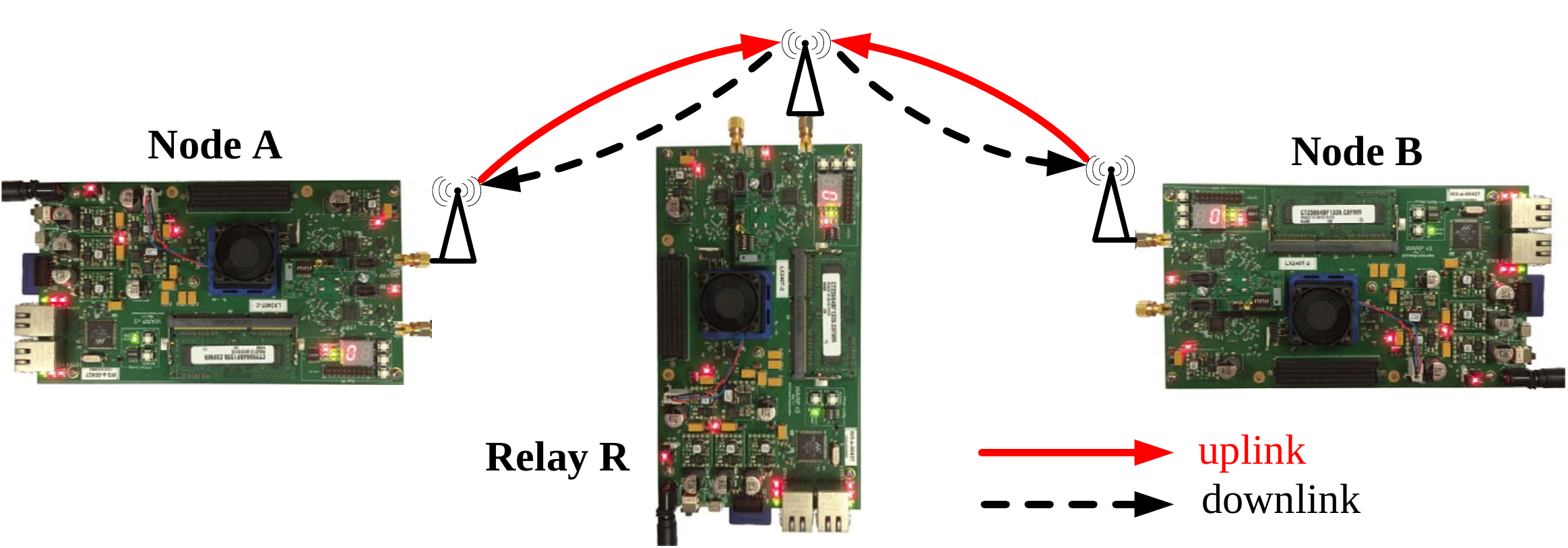}
		\par\end{centering}
	
	\caption{\label{fig:PNC-WARP}PNC on WARP. }
	
\end{figure}

Fig. \ref{fig:PNC-WARP} shows our SDR implementation of PNC on the wireless open access research platform (WARP) \cite{warpproject}. In our experimental set-up, three WARP nodes serve as the relay, node A, and node B. In what follows, we put together all the functionalities described in the preceding sections for an overall design, and describe the FPGA implementation of time-critical functions that achieves an effective feedback delay (from channel estimation to using the channel estimate for precoding) of about $0.5ms$.

\begin{figure*}[t]
	\begin{centering}
		\subfloat[\label{fig:Diagram-RX}Processing of the preambles and postambles at the receiver of the relay.]{\begin{centering}
				\includegraphics[width=0.89\textwidth]{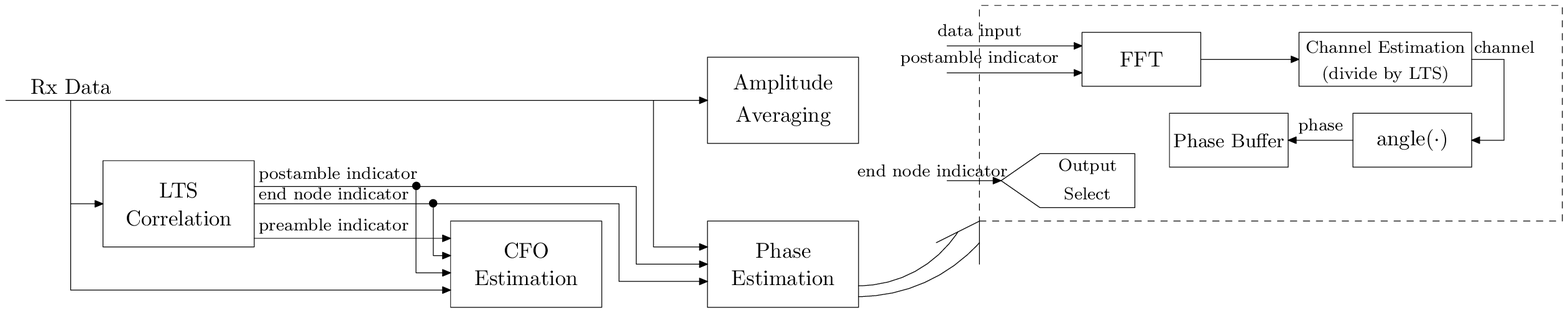}
				\par\end{centering}
			
		}
		\par\end{centering}
	
	\begin{centering}
		\subfloat[\label{fig:Diagram-TX}Transmitter of the end nodes.]{\begin{centering}
				\includegraphics[width=0.98\textwidth]{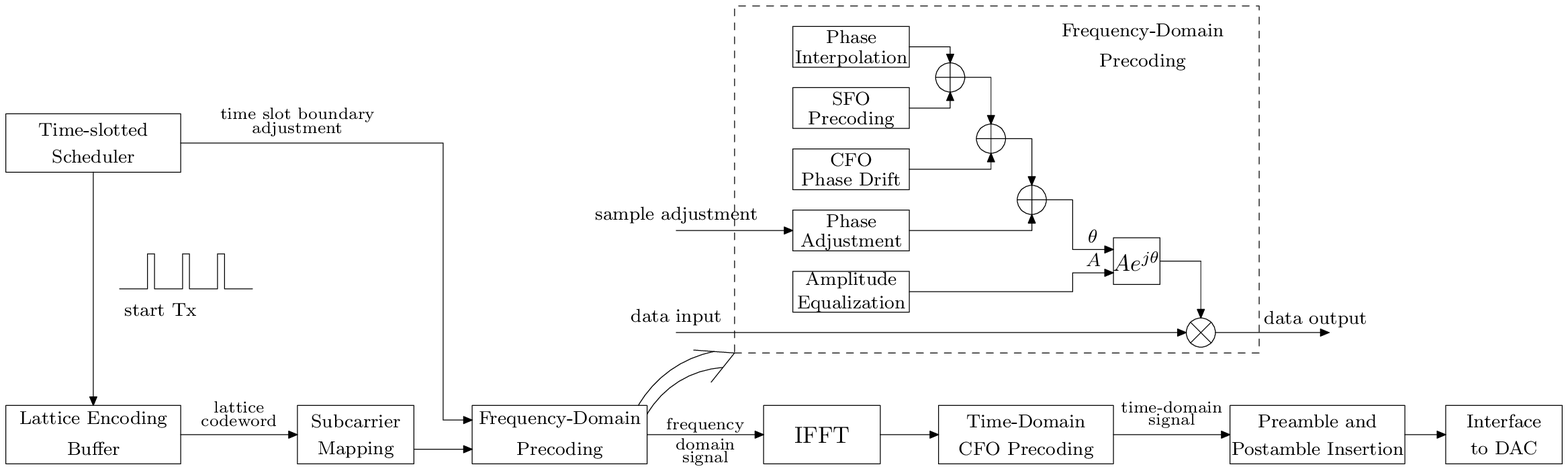}
				\par\end{centering}
			
		}
		\par\end{centering}
	
	\caption{\label{fig:System-Diagrams}System Diagrams.}
	
\end{figure*}

Fig. \ref{fig:System-Diagrams} shows the block diagrams of the FPGA receiver design of the relay and the transmitter design of the end nodes, based on WARP 802.11 reference design. The receiver of the end nodes and the transmitter of the relay follow the standard of the OFDM PHY layer of IEEE 802.11, but without the MAC layer of IEEE 802.11.

In the receiver diagram of the relay, the LTS correlation block determines the sample positions of the preambles and postambles in the received data, and indicates whether the preamble or postamble currently under processing belongs to node A or B. The CFO estimation block estimates the CFO of each node based on \eqref{eq:CFO-estimation} and the rough CFO. The channel estimation block estimates the frequency-domain (subcarrier) channels of each end node based on the received postamble. The amplitude averaging block averages the amplitudes of pilots to get an overall amplitude scaling factor for feedback to the end node. All the above processing can be completed within $10\mu s$ after the reception of the postamble. 

In the transmitter diagram of the end nodes, the time-slotted scheduler schedules packet transmissions according to the timer that was set in the initialization phase. It also adjusts the timer occasionally (when requested by the relay) to align slot boundaries. Each packet contains five lattice codewords. The lattice codewords are first mapped to the OFDM subcarriers. Then the frequency-domain precoding block precodes both the phase and amplitude for different subcarriers. In the frequency-domain precoding block, the phase interpolation block recovers the phase of each subcarrier by interpolating with the partial feedback of phases. The CFO phase drift block calculates the phase drift due to the CFO from the time when the postambles of last packet was transmitted to the time when the data part of the current packet will be transmitted. The SFO precoding block calculates the SFO phase shift from the time when the postamble of last packet was transmitted to the time when the current OFDM symbol of the current packet will be transmitted. The phase adjustment block implements \eqref{eq:Phase-adjustment}. The amplitude equalization block calculates the absolute amplitude precoding factors for the uplink subcarriers based on the relative downlink amplitudes of the subcarriers and the feedbacked amplitude scaling factor. The amplitude precoding factor and the phase of each subcarrier then form a complex precoding factor for the subcarrier that will be multiplied to the frequency-domain signal before transmission. After that, the time-domain CFO precoding block compensates for the phase drift in the data part. All the above precoding processing can be completed within $10\mu s$ after the reception of the downlink packet from the relay. That is, the transmission of the next uplink packet can be started within $10\mu s$ of the reception of the downlink packet.

\section{Experiments}\label{sec:Experiments}
This section presents our experimental results. In our experiments, the WARP nodes are placed in an indoor office environment, with pairwise separation ranging from $2$ meters to more than $10$ meters. The carrier frequency is $2.5GHz$. The bandwidth (and the sampling rate) is $20MHz$. The CFO between each pair of WARP nodes range from $5kHz$ to $10kHz$, and can change by hundreds of hertz as shown by the experiment in Fig. \ref{fig:CFO-variation}. The time slot duration of PNC is $T_{\mathrm{slot}}=1ms$, with $0.5ms$ for uplink followed by $0.5ms$ for downlink. 

The measurements of our phase precoding has already been shown in Figs. \ref{fig:Constellations-aligned} and \ref{fig:Empirical-CDFs}. To demonstrate the advantage of PNC over traditional methods and to justify the motivation of PNC, in this section we compare the BER of TD, SNC, and PNC in two-user relay networks. We also present the symbol error rate (SER) of PNC with different modulations. Finally, we present the throughputs of lattice-coded PNC with different code rates, in both static LoS scenario and mobile non-LoS scenario, and we also show that PNC can achieve higher throughput compared with SNC.

\subsection{BER of TD, SNC, and PNC}\label{sub:ber-compare}
To compare TD, SNC, and PNC fairly, we did an experiment to measure the BER of the three schemes for the same normalized end-to-end data delivery rate with QAM modulation without channel coding. As TD, SNC, and PNC take $4$, $3$, and $2$ non-overlapping transmissions respectively, we use 256-QAM ($8$ bits/symbol) for TD, 64-QAM ($6$ bits/symbol) for SNC, and 16-QAM ($4$ bits/symbol) for PNC, such that they have the same normalized end-to-end data delivery rate of  $\frac{8}{4}=\frac{6}{3}=\frac{4}{2}=2$ bits/symbol. We do not incorporate channel coding so as to have a clean comparison of the transmission schemes. 

\begin{figure}[h]
	%\vspace{-0.5em}
	\begin{centering}
		\includegraphics[width=0.7\columnwidth]{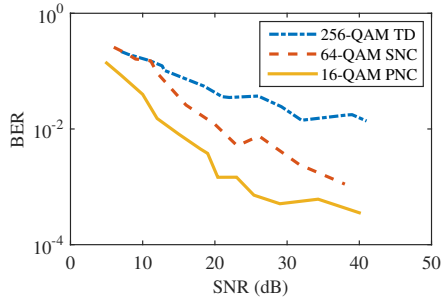}
		\par\end{centering}

	\caption{\label{fig:ber-compare}BER of uncoded TD, SNC, and PNC for the same
		normalized end-to-end data delivery rate. }
	\vspace{-0.5em}
\end{figure}

The experiment results in Fig. \ref{fig:ber-compare} demonstrate that PNC can achieve lower BER for the same normalized data delivery rate, especially at low-SNR regime. This is because PNC can afford to use the lower-order modulation for the same data delivery rate. 

\subsection{SER of PNC with Different Modulations}
\begin{figure}[h]
	\vspace{-0.5em}
	\begin{centering}
		\includegraphics[width=0.7\columnwidth]{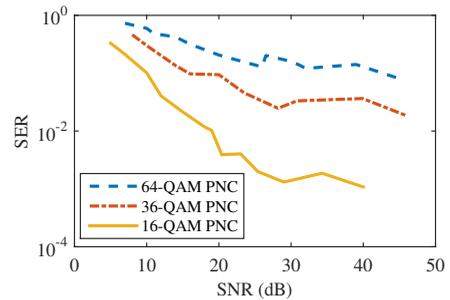}
		\par\end{centering}
	\vspace{-0.5em}
	\caption{\label{fig:ser-compare-mod}SER of PNC with different QAM modulations. }
\end{figure}

Fig. \ref{fig:ser-compare-mod} shows the SER of PNC with different QAM orders. Higher-order modulations yield higher data rates, but the SER will also be higher. Moreover, the SER curves of 36-QAM and 64-QAM have SER floors, beyond which the SER cannot be further reduced by increasing the SNR. The SER floors are due to the precoding errors caused by various factors, such as channel estimation errors, frequency offset estimation errors, channel variations, and oscillator phase noise. Increasing SNR cannot deal with the errors of all these factors; hence the error floors. Higher-order modulations are more susceptible to the precoding errors, and have higher error floors.

\subsection{Throughput Performance of Lattice-Coded PNC}
We now present the throughput performance of the overall lattice-coded PNC system in static LoS scenario and mobile non-LoS scenario. Each LDLC codeword of $n=960$ symbols is mapped to $N_s=20$ OFDM symbols, each consisting of the $48$ data-carrying OFDM subcarriers (i.e., the data subcarriers in 802.11). Each uplink packet contains $N_c=5$ LDLC codewords, i.e. $100$ OFDM symbols, the preamble, and the postamble. The source message associated with each LDLC codeword consists of m complex source symbols. The real and complex parts of each source symbol are elements of the integer alphabet $\mathcal{A}=\left\{-1,0,1\right\}$. Taking into account the overhead of the preamble and postamble, the silence period, and the downlink period, the maximum throughput (if all messages are delivered successfully) of each end node is $r_{n}=\frac{N_{c}\cdot m\cdot\log_{2}9}{T_{\mathrm{slot}}}$. In our experiments below, we evaluate the throughput using different code rates $\frac{m}{n}$ by setting $m=800$ ($r_n=12.7Mbps$) and $m=900$ ($r_n=14.3Mbps$). For the mobile non-LoS experiments, the end nodes were randomly placed at locations where the direct path to the relay was blocked by obstacles. Two persons held the two end nodes separately and walked at ordinary speed.

\begin{figure}[h]
	\vspace{-0.5em}
	\begin{centering}
		\includegraphics[width=0.7\columnwidth]{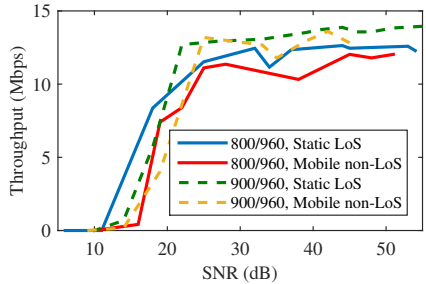}
		\par\end{centering}
	\vspace{-0.5em}
	
	\caption{\label{fig:Throughput}Normalized PNC throughput with different code rates in static LoS and mobile non-LoS scenarios. }
\end{figure}

As shown in Fig. \ref{fig:Throughput}, our lattice-coded PNC system achieves good throughput performance using code rates $800/960$ and $900/960$ in the static LoS and the mobile non-LoS scenarios. In the low SNR regime, the mobile non-LoS scenario has a penalty of $2\sim4dB$ with respect to the static LoS scenario. In the high SNR regime, our system can approach the maximum throughput in the static LoS scenario and have a small gap from the maximum throughput in the mobile non-LoS scenario. The penalties in the mobile non-LoS scenario are incurred by the channel variation due to motion and frequency selective fading (probably deep fading in some subcarriers). Fig. \ref{fig:Throughput} also shows that code rate $900/960$ yields higher throughput in the high SNR regime than that of code rate $800/960$, while code rate $800/960$ yields higher throughput in the low SNR regime since lower code rate can reduce packet error rate. We remark that without channel alignment, the throughputs in all the above cases will be practically zero. 

\begin{figure}[h]
	\vspace{-0.5em}
	\begin{centering}
		\includegraphics[width=0.7\columnwidth]{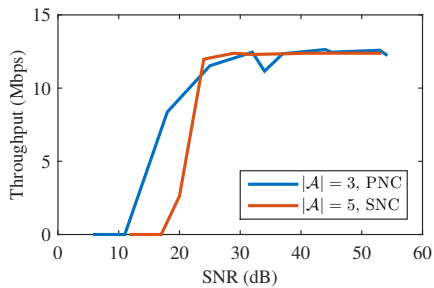}
		\par\end{centering}
	\vspace{-0.5em}
	
	\caption{\label{fig:Throughput-compare}Throughputs of PNC and SNC. }
\end{figure}

We compare the throughputs of lattice-coded PNC with alphabet $\mathcal{A}=\left\{-1,0,1\right\}$ and SNC with alphabet $\mathcal{A}_1=\left\{-2,-1,0,1,2\right\}$, both with coding rate of $800/960$, in static LoS scenarios; Fig. \ref{fig:Throughput-compare} shows the results. If no packet errors occur, the throughput of the two schemes will be very close, since $\frac{\log\left(25\right)}{3}\approx \frac{\log\left(9\right)}{2}$. But as PNC can use a smaller alphabet, PNC has lower packet error rate and thus higher throughput in the moderate SNR regime. 

\subsection{Comparison with Prior PNC Implementations}
\label{sub:prior-pnc}

\begin{table*}[th]
	\caption{\label{tab:pnc}PNC implementations.}
	\begin{centering}
		\begin{tabular}{|>{\centering}p{1cm}|c|>{\centering}p{1.6cm}|c|>{\centering}p{2.3cm}|>{\centering}p{1.8cm}|}
			\hline 
			& Network & Modulation & Bandwidth & Channel Coding & Note\tabularnewline
			\hline 
			\cite{lu2013implementation} & TWRN & BPSK & $4MHz$ & Convolutional & \tabularnewline
			\hline 
			\cite{chen2013frequency} & TWRN & BPSK & $0.019MHz$ & Uncoded & \tabularnewline
			\hline 
			\cite{lu2014network} & NCMA & BPSK & $4MHz$ & Convolutional & \tabularnewline
			\hline 
			\cite{wu2014analysis} & TWRN & QPSK & $0.35MHz$ & LDPC & High BER\tabularnewline
			\hline 
			\cite{yang2015asynchronous} & TWRN & BPSK & $2MHz$ & Convolutional & \tabularnewline
			\hline 
			\cite{youncmaii2015} & NCMA & BPSK & $5MHz$ & Convolutional & \tabularnewline
			\hline 
			\cite{kramarev2015implementation} & TWRN & QPSK & $3.125MHz$ & RS code & GPSDO\tabularnewline
			\hline 
			\cite{marcum2015analysis} & TWRN & BPSK & $2MHz$ & - & \tabularnewline
			\hline 
			\cite{pan2015network} & NCMA & QPSK & $5MHz$ & Convolutional & Two antennas\tabularnewline
			\hline 
			\cite{you2016reliable} & TWRN & BPSK & $5MHz$ & Convolutional & \tabularnewline
			\hline 
			This paper & TWRN & 16-QAM, 9-QAM & $20MHz$ & Lattice codes (LDLC) & \tabularnewline
			\hline 
		\end{tabular}
		\par\end{centering}
\end{table*}

There were several PNC implementations prior to our work \cite{lu2013implementation, chen2013frequency, lu2014network, wu2014analysis, yang2015asynchronous, youncmaii2015, kramarev2015implementation, marcum2015analysis, pan2015network, you2016reliable}. These implementations did not use channel precoding. Tab. \ref{tab:pnc} compares the modulation, bandwidth, and channel coding of these implementations with ours. Ref. \cite{lu2013implementation} implemented TWRN with BPSK; it was the first PNC implementation. Ref. \cite{chen2013frequency} investigated receiver-side CFO compensation for BPSK signals, and implemented TWRN with differential BPSK (DBPSK). Ref. \cite {lu2014network} implemented a network-coded multiple access (NCMA) network with BPSK. Ref. \cite{wu2014analysis} implemented TWRN with QPSK, but the demonstrated BER (with LDPC coding) could be more than $0.1$ for CFO less than $1kHz$. Ref. \cite{yang2015asynchronous} dealt with integral and fractional symbol misalignments, and implemented TWRN with BPSK. Ref. \cite{youncmaii2015} implemented a real-time NCMA network with BPSK. Ref. \cite{kramarev2015implementation} implemented TWRN with QPSK, aided by highly accurate and expensive GPSDO to ensure that the phase of the baseband channel only varies slowly. Ref. \cite{marcum2015analysis} implemented TWRN with BPSK. Ref. \cite{pan2015network} implemented NCMA with QPSK using two antennas at the receiver. Ref. \cite{you2016reliable} implemented a TWRN system that can support real TCP/IP applications such as video conferencing. These implementations used bandwidth of ${<}5MHz$; higher bandwidth requires faster processing (which is a challenge to SDR development using universal software radio peripheral (USRP) and GNUradio) and tighter synchronization. The implementation in \cite{kramarev2015implementation} used the RS code, and those in \cite{lu2013implementation, lu2014network, yang2015asynchronous, youncmaii2015, pan2015network, you2016reliable} used convolutional codes. In comparison to these implementations, our work can support higher-order modulations (16-QAM and 9-QAM), higher bandwidth ($20MHz$), and lattice codes, without the need for expensive oscillators or multiple antennas.

\section{Discussions}\label{sec:discussions}
Though this paper considers the TWRN setup, much of the discussion and the proposed system design also applies to CF networks, where multiple source nodes transmit to a destination node via multiple relay nodes. The discussions on LDLC in Section \ref{sec:LDLC} apply to CF networks in a straightforward manner. To achieve time-slot synchronization in CF networks, we can let the source nodes synchronize to a reference relay using the method proposed in Section \ref{sub:Time-Slotted-System}. Then the signals of the source nodes will also be synchronized at the other relays, with only small constant time offsets that are different from that with respect to the reference relay (because of relative different path lengths from sources to relays). Furthermore, the source nodes can precode for the frequency offsets (CFO and SFO) such that they will have common frequency offsets with respect to each relay. Then each relay can compensate for the common frequency offsets to remove the phase drifts throughout a packet, to facilitate the computation of a combination with fixed integer coefficients. For this purpose, it is also important to estimate the CFO accurately using the method in Section \ref{sub:Channel-Estimation}. The channel precoding in CF networks is rather different from that in TWRN, because the channels of the source nodes can neither be amplitude-aligned nor phase-aligned to multiple relays at the same time in general (e.g., the aforementioned constant time offsets introduce different phases in the OFDM domain). Nevertheless, the reciprocity-based channel estimation (of both amplitude and phase) in Section \ref{sub:Amplitude-Precoding}, \ref{sub:Channel-Estimation}, and \ref{sub:reciprocal-phase} still works in CF networks, and the source nodes can align their channels to a selected relay. If full-feedback channel information is available at a centralized node, it can jointly optimize the precoding coefficients of the source nodes to make it relatively easy for all relays to compute integer-coefficient combinations. As the channels cannot be aligned to multiple relays simultaneously, different subcarriers may have different channel coefficients, and thus different optimal integer coefficients for combination computing (in contrast to the all-one coefficients in channel-aligned TWRN). If a relay uses the same integer coefficients to compute the combinations on all subcarriers, the decoding error rate may be high. If a relay uses different integer coefficients to match the channels of different subcarriers better, the different integer coefficients need to be delivered to the destination together with the message combination as meta data. The total overhead of the integer coefficients for all relay nodes is $O\left(L\times M\times N_d \right)$, where $L$ is the number of source nodes, $M$ is the number of relay nodes, and $N_d$ is the number of data subcarriers. Moreover, when different integer coefficients are used for different subcarriers, a codeword should be placed on a single subcarrier (or multiple subcarriers that have the same integer coefficient) only, because otherwise the combination will not be a valid codeword. So, $N_d$ different subcarriers will be used to convey $N_d$ different codewords, with the codeword length limited by the number of OFDM symbols.

Our channel precoding system is also useful to distributed MIMO systems. In distributed MIMO, multiple APs transmit to multiple client nodes. The clients are distributed, and the APs have a backhaul connection (typically Ethernet). The APs do zero-forcing beamforming (ZFBF) jointly to invert the MIMO channel and let each client only receive its desired signal. To enable the ZFBF at the APs, we may need the clients to feedback their channel information \cite{balan2013airsync}. The APs can also use a relative calibration technique \cite{rogalin2014scalable} to estimate the downlink channel based on reciprocity, so as to remove the need for feedback for the support of legacy client devices that have no feedback capability. The relative calibration requires the APs to share their channel information through the backhaul network. Both methods induce communication overhead of channel information, either in the wireless network or in the backhaul network. We can reduce the overhead greatly using our partial feedback channel estimation described in Section \ref{sec:Channel-Alignment}. 

It may be also interesting to consider integrating PNC with interference alignment \cite{gollakota2009interference}. Using interference alignment, we may align the undesired interference signals from multiple transmitters to the same direction in the space domain. Using the idea of PNC, the receiver may decode a combination from the overlapped undesired interference signals. Then the receiver can forward the undesired signal combination to another receiver that desires the combination. 

In our experiment, the distance between WARP nodes was shorter than $10$ meters, and the propagation delay for $10$ meters is just less than one sample for $20MHz$ bandwidth. When the distance between A and R and the distance between B and R differ by $300$ meters, the difference of propagation delays will differ by $20$ samples. Then, after the initial beacon triggering (which triggers the end nodes to set their timers), the uplink packets of the two end nodes will miss by the round-trip delay difference, i.e. $40$ samples. Since it is larger than the CP length of the LTS, it will destroy the orthogonality of the LTS of the two end nodes, and the relay may fail to find the time offsets of the two end nodes correctly. Then the uplink data transmissions in the later time slots may still suffer large time offsets and the relay cannot decode correctly.  To ensure the orthogonality of the LTS of the two end nodes, we can lengthen the CP or estimate the propagation delays and let the end nodes adjust their time slot boundaries accordingly. We can estimate round-trip propagation delays first, and then divide that by two to get the single-trip propagation delays. 

Our channel precoding system is designed for indoor environment, where wireless nodes are static or moving slowly. In vehicular networks, the relative movement speed between two nodes can be as high as $50m/s$, translating to relative movement length of $5cm$ in $1ms$. Since the wavelength is ($\frac{3\times10^{8}m/s}{2.5\times10^{9}Hz}=12cm$ for carrier frequency of $2.5GHz$ ($5.1cm$ for $5.9GHz$), the movement will induce significant phase changes, invalidating the channel precoding.

\section{Conclusion}\label{sec:Conclusions} 

This paper presented a first implementation of a practical lattice-coded PNC system. Our system achieves accurate channel alignment of distributed nodes using only temperature-compensated oscillators, without the need for extra antennas or bandwidth to broadcast reference signal. The accurate channel alignment is attributed to two critical components: 1) a CFO estimation method that is $100$ times more accurate than the conventional 802.11 method; 2) fast CSI feedback in the ballpark of $0.5ms$. Our channel precoding system also has very low signaling overhead and, by exploiting channel reciprocity, it requires partial CSI feedback that is only as little as $1\%$ of the data payload. In addition to the tight channel alignment, we also redesigned the lattice encoding and decoding algorithms to deal with the challenges of encoding/decoding complexity and residual channel misalignments. We demonstrated good throughput performance of the implemented PNC system in static LoS and mobile non-LoS scenarios. Prior to our work here, investigations of lattice-coded communication systems (PNC or non-PNC) existed primarily in the theoretical domain. We believe our work is an important first step toward making lattice-coded communication systems practical.

%\IEEEpubidadjcol

% or


%\appendices

\ifCLASSOPTIONcompsoc
  %\section*{Acknowledgments}
\else
  %\section*{Acknowledgment}
\fi


\ifCLASSOPTIONcaptionsoff
  %\newpage
\fi

\end{document}